\documentclass[12pt]{article}
\usepackage{amsmath,amssymb}
\usepackage[dvips]{graphicx}
\renewcommand{\theequation}{\thesection.\arabic{equation}}
\renewcommand{\thefootnote}{\fnsymbol{footnote}}
\newcommand{\EQ}{\begin{equation}}
\newcommand{\EN}{\end{equation}}
\newcommand{\bea}{\begin{eqnarray}}
\newcommand{\ena}{\end{eqnarray}}
\newcommand{\vs}[1]{\vspace{#1 mm}}

\newcommand{\uda}{\nearrow \kern-1em \searrow}
\makeatletter
\def\eqnarray{%
 \stepcounter{equation}%
 \let\@currentlabel=\theequation
 \global\@eqnswtrue
 \global\@eqcnt\z@
 \tabskip\@centering
 \let\\=\@eqncr
 $$\halign to \displaywidth\bgroup\@eqnsel\hskip\@centering
 $\displaystyle\tabskip\z@{##}$&\global\@eqcnt\@ne
 \hfil$\displaystyle{{}##{}}$\hfil
 &\global\@eqcnt\tw@$\displaystyle\tabskip\z@{##}$\hfil
 \tabskip\@centering&\llap{##}\tabskip\z@\cr}
\makeatother
\def\p{\partial}
\def\pr{\prime}

\def\t{\tilde}\def\wt{\widetilde}

\def\cal{\mathcal}\def\wh{\widehat}
\def\le{\left}\def\ri{\right}
\def\fr{\frac}
\def\t{\tilde}\def\wt{\widetilde}
\def\nonum{\nonumber}
\usepackage{color}
\definecolor{yuta}{cmyk}{0.85,0.90,0.33,0.21}
\definecolor{Sepia}{cmyk}{0.00,0.83,0.44,0.70}
\definecolor{BlueGreen}{cmyk}{0.85,0.00,0.33,0.00}
\definecolor{VioletRed}{cmyk}{0.94,0.54,0.00,0.30}
\definecolor{yutaRed}{cmyk}{0.00,1.00,0.90,0.10}
\definecolor{Gray}{cmyk}{0.00,0.00,0.00,0.50}
\definecolor{Plum}{cmyk}{0.50,1.00,0.00,0.00}
\definecolor{VioletRed}{cmyk}{0.65,0.81,0.00,0.00}
\definecolor{RedViolet}{cmyk}{0.07,0.90,0.00,0.34}
\definecolor{BlueViolet}{cmyk}{0.86,0.91,0.00,0.4}
\definecolor{GreenYellow}{cmyk}{0.15,0.00,0.69,0.00}
\definecolor{Cyan}{cmyk}{1.00,0.00,0.69,0.00}
\definecolor{SkyBlue}{cmyk}{0.62,0.00,0.12,0.00}
\definecolor{PineGreen}{cmyk}{0.92,0.00,0.59,0.25}
\definecolor{Emerald}{cmyk}{1.00,0.00,0.50,0.00}
\definecolor{Aquamarine}{cmyk}{0.82,0.00,0.30,0.00}
\definecolor{Turquoise}{cmyk}{0.85,0.00,0.20,0.00}
\definecolor{RedOrange}{cmyk}{0.00,0.77,0.87,0.00}
\definecolor{RedOrange}{cmyk}{0.00,0.77,0.87,0.00}
\definecolor{Melon}{cmyk}{0.00,0.46,0.50,0.00}
\definecolor{Apricot}{cmyk}{0.00,0.32,0.52,0.00}
\definecolor{Goldenrod}{cmyk}{0.00,0.10,0.84,0.00}
\definecolor{Lavender}{cmyk}{0.00,0.48,0.00,0.00}
\definecolor{Orchid}{cmyk}{0.32,0.64,0.00,0.00}
\definecolor{Thistle}{cmyk}{0.12,0.59,0.00,0.00}
\definecolor{SeaGreen}{cmyk}{0.69,0.00,0.50,0.00}
\definecolor{Red}{cmyk}{0.00,1.00,1.00,0.00}
\definecolor{Blue}{cmyk}{1.00,1.00,0.00,0.00}
\definecolor{Green}{cmyk}{1.00,0.00,1.00,0.00}
\definecolor{Dandelion}{cmyk}{0.00,0.29,0.84,0.00}
\definecolor{Peach}{cmyk}{0.00,0.50,0.70,0.00}

\definecolor{Green2}{cmyk}{0.8,0.2,0.8,0.5}

\begin{document}

\begin{titlepage}
\setcounter{page}{0}
\begin{flushright}
EPHOU 10-004\\
\today\\
\end{flushright}

\vs{6}
\begin{center}
{\Large \textbf{A Note on Computations of D-brane Superpotential}}

\vs{6}
\textbf{Hiroyuki Fuji}\\
{\small \em Department of Physics, Nagoya University,\\
Nagoya 466-8602, Japan}\\
\textbf{Shinsaku Nakayama, \ Masahide Shimizu \ and \ Hisao Suzuki}\\
{\small \em Department of Physics, 
Hokkaido University, \\Sapporo 060-0810, Japan } \\
\end{center}
\vs{6}

\centerline{{\bf{Abstract}}}
We develop some computational methods for the integrals over the $3$-chains 
on the compact Calabi-Yau $3$-folds
that plays a prominent role in the analysis of the topological B-model 
in the context of the open mirror symmetry.
We discuss such $3$-chain integrals in two approaches. In the first approach, 
we provide a systematic algorithm to obtain the inhomogeneous Picard-Fuchs 
equations. In the second approach, we discuss the analytic continuation 
of the period integral to compute the $3$-chain integral directly. 
The latter direct integration method is applicable for both on-shell and
 off-shell formalisms. 
\end{titlepage}
\newpage

\renewcommand{\thefootnote}{\arabic{footnote}}
\setcounter{footnote}{0}

\tableofcontents
\section{Introduction}
The progress of the research on mirror symmetry has been 
active remarkably in recent years. 
The mirror symmetry is a powerful tool to study the non-perturbative 
aspects of effective field theory arising from type II string compactifications. 
In the recent developments, the mirror symmetry of the open string sector has become tractable, 
and it is possible to compute the effective superpotential on D-brane, which wraps around the cycles in Calabi-Yau $3$-fold. 
The concrete study of the mirror symmetry for the open string sector was initiated in \cite{AV}, 
and the enumerative structure predicted in \cite{OV} is confirmed on the toric Calabi-Yau case. 
These results are confirmed by the consistency with the topological vertex computations \cite{AKMV}. 

Recently, the open mirror symmetry has been extended to certain {\it compact} Calabi-Yau manifolds. 
Walcher \cite{Wa1} predicted the disc instanton numbers 
for the quintic $3$-fold with an involution brane 
as a natural extension of the closed mirror symmetry \cite{CDGP}. 
This was proven rigorously by localization calculation 
on the A-model side in \cite{PaSoWa}. 
Then the B-model analysis is precisely developed in \cite{MW}. 
In particular for the B-model side, the computation of the $3$-chain integrals, which is obtained by the reduction of the holomorphic Chern-Simons action, 
plays a prominent role. 
They are the solutions of the Picard-Fuchs equation with a inhomogeneous term resulting from the boundary contribution. 
The open mirror symmetry with an involution brane 
for the other compact Calabi-Yau $3$-folds are further studied in \cite{KW1,KS1,Walcher}. 

The relative period for compact Calabi-Yau is studied in \cite{JS1} 
by replacing curves with a divisor and adding logarithmic factor in the period. 
Relating to this work, the study of the toric branes in the compact Calabi-Yau geometry is also developed in \cite{AHMM} 
along a similar line as \cite{AV}. 
There are some related works \cite{GHKK,JS2,AHJMMS,GHKK2,AB,worldsheet,GKZ,LLY}. 
The toric brane is specified by the extra toric charges, and an open string moduli is introduced. 
The effective superpotential on the toric brane is given 
by the relative period integrals, which are the integral of holomorphic $3$-form over $3$-chains with the boundaries. 
The relative period satisfies the extended Picard-Fuchs equation \cite{LM,LMW1,LMW2} which depends on both closed and open string moduli. 
See also \cite{Lerche} for review. 
Extremizing the effective superpotential with respect to the open moduli, one finds the same results as the involution brane \cite{JS1,AHMM}. 
Therefore the effective superpotential for the involution brane is called as {\it on-shell}.

The aim of this article is to study the B-model side of open mirror symmetry. 
First, we will discuss the computation of the inhomogeneous term \cite{MW,LLY} in 
the Picard-Fuchs equation for the $3$-chain integral associated to 
the holomorphic curves which is mirror to the involution brane. 
The inhomogeneous Picard-Fuchs equation is derived by performing the 
Griffiths-Dwork algorithm \cite{Griffiths}. 
The algorithm itself is clear and straightforward, but the explicit computation
needs some efforts. In this article, we will propose a more efficient
computational method for the Griffiths-Dwork algorithm by considering the 
$3$-chain integral more precisely. 
Taking the integration by parts for the $3$-chain integral successively, 
we find a ring structure which is generated by the boundary terms.
The ring structure makes the Griffiths-Dwork algorithm more manifest,
and one can obtain the inhomogeneous Picard-Fuchs equation rather efficiently.

Second, we will discuss the direct integration of the $3$-chain integral. 
In the study of the original mirror symmetry \cite{CDGP}, 
the period integral is computed by the direct 
integration of the holomorphic $3$-form over the $3$-cycle. 
In general, these periods are obtained systematically as the solutions of the Picard-Fuchs equation \cite{KT1}, 
but the direct computation is still meaningful because of its clear geometric picture. 
In this article, we will consider the direct integration of the 
$3$-chain integral via analytic continuation. In \cite{JS2} a similar 
computation is discussed for the relative periods. 
The main point of our work is the direct evaluation of 
the $3$-chain integral itself without computing the other periods 
which form the Gauss-Manin system for the ${\cal N}=1$ special geometry.\footnote{
The original framework of $\mathcal{N}=1$ special geometry is discussed in \cite{LMW1,LMW2,mayr} and 
the application to the compact Calabi-Yau is discussed in \cite{JS1}. 
} 

In the analytic continuation, we replace the formal power sum 
with respect to the complex structure moduli in the period integral by the residue integrals. 
In the evaluation of the fundamental period, the poles in the residue
integrals only appear at the integral values, and we obtain 
the power sum solution of the fundamental period around the 
large complex structure point. 
For $3$-chain integral, the poles at half-integer points appear 
in the integrand of the residue integral. 
Picking up all the half-odd points in the residue integrals, we obtain 
the celebrated solution for the superpotential on the involution brane. 
Our method is advantageous for the computations of the Calabi-Yau 
$3$-folds with multiple moduli, because in these cases the inhomogeneous Picard-Fuchs 
equation becomes too complicated. 

Finally, we extend our method for the direct computation to the 
evaluation of the relative period integrals which appear in a so-called
{\it off-shell  formalism}. 
The Poincar\'e residue theorem implies that the relative period integral 
can also be computed by introducing the logarithmic factor which restricts 
the integral to the divisor locus \cite{JS1}. 
Therefore we are able to apply our method to the integral of the holomorphic $3$-form with a logarithmic factor. 
The results of our computation coincide with those of \cite{JS1,AHMM,AB}, 
and we can check that they satisfy the extended Picard-Fuchs equation 
for the toric brane. 
We can also verify that at the critical locus of the open moduli the relative period coincides the on-shell results. 

This paper is organized as follows. 
In section $2$, we discuss the ring structure of the Griffiths-Dwork algorithm.
The inhomogeneous terms in the Picard-Fuchs equation are computed explicitly for 
some one-parameter complete intersection models. 
These examples are already considered in \cite{Wa1,Wa2,Walcher}
and our computation recovers correctly their results.
In section $3$, we discuss the direct computation of the $3$-chain
integral via analytic continuation. We will mainly focus on the two basic 
examples, quintic hypersurface in $\mathbb{CP}^4$ and double cubic complete intersection in $\mathbb{CP}^5$, 
and find the solutions of the inhomogeneous Picard-Fuchs equations. 
We also consider one of the two-parameters examples considered in \cite{Walcher} and check that its result can be reproduced easily. 
In section $4$, we extend our analysis of the direct computation to the relative period integrals. 
We first discuss how the integral of the holomorphic $3$-form with 
a logarithmic factor yields to the similar form as  the $3$-chain 
integral, and then we evaluate the integrals for the above two one-parameter models. 
In section $5$ we try to fix the normalization ambiguities resulted from analytic continuations. 
In section $6$, we comment our conclusions. 
In the appendix A, we present the details of computations in section $2$. 

\section{Inhomogeneous Picard-Fuchs equations}
\subsection{An alternative technique to Griffiths' reduction}
In order to consider the chain integrals of holomorphic $3$-forms and derive the Picard-Fuchs equations, 
there is a useful method known as the Griffiths-Dwork method.
We will introduce here this method briefly adopting the mirror quintic as an example.
The mirror quintic Calabi-Yau $3$-fold $Y_5$ is defined by the degree five homogeneous polynomial in $\mathbb{CP}^{4}/{(\mathbb{Z}_5)^3}$, 
 \begin{eqnarray}
  W=\frac{1}{5}\left({x}_{1}^{5}+x_{2}^{5}+x_{3}^{5}+x_{4}^{5}+x_{5}^{5}\right) -\psi x_{1}x_{2}x_{3}x_{4}x_{5}=0 \label{quintic}
 \end{eqnarray}
 where $x_{1}$, $x_{2}$, $\cdots$, $x_{5}$ are the homogeneous coordinates of $\mathbb{CP}^{4}$ and $\psi$ is a parameter, 
which is a complex structure moduli, 
of the hypersurface. 
The holomorphic $3$-form $\Omega(z)$ is given as the residue 
at the loci $W=0$ on the ambient $\mathbb{CP}^{4}$ by
 \begin{eqnarray}
  \Omega(z)=\mathrm{Res}_{W=0} \wt{\Omega}(z), \quad \wt{\Omega}(z)=\psi\fr{\omega_{0}}{W},
 \end{eqnarray}
 where $z=(5\psi)^{-5}$. 
The holomorphic $4$-form $\omega_{0}$ 
is defined by
 \begin{eqnarray}
  \omega_{0}=\sum_{i=1}^{5}(-1)^{i-1}x_{i}dx_{1}\wedge\cdots\wedge\wh{dx}_{i}\wedge\cdots\wedge dx_{5},
 \end{eqnarray}
where $\wh{dx}_{i}$ implies the absence of $dx_{i}$ in the summation.
For the small tubular neighborhood $T_{\epsilon}(\Gamma)$ around $\Gamma$, one can express the integral of the holomorphic $3$-form  
 \begin{eqnarray}
  \int_{\Gamma}\Omega(z)=\fr{1}{2\pi i}\int_{T_{\epsilon}(\Gamma)}\wt{\Omega}
  =\fr{\psi}{2\pi i}\int_{T_{\epsilon}(\Gamma)}\fr{\omega_{0}}{W}. \label{G_Dwork}
 \end{eqnarray}
To derive the Picard-Fuchs equation, 
the systematic algorithm proposed by Griffiths 
is applied for the reduction of the pole order in $\tilde{\Omega}$. 
In this paper, instead of adopting the Griffiths' reduction algorithm, we
will develop an alternative way to derive inhomogeneous Picard-Fuchs
equation. 
Our method is an extension of the procedure which is studied for the 
period integrals without boundaries \cite{Mo,HKTY,KLRY}. 

We will consider the mirror quintic defined $\eqref{quintic}$ as an example. 
Let us represent the defining polynomial $\eqref{quintic}$ with the redundant coefficient parameters $a_{i}$ of each monomial: 
 \begin{eqnarray}
  W'=a_{1}x_{1}^{5}+a_{2}x_{2}^{5}+a_{3}x_{3}^{5}+a_{4}x_{4}^{5}+a_{5}x_{5}^{5}+a_{0}x_{1}x_{2}x_{3}x_{4}x_{5}. \label{quintic2} 
 \end{eqnarray} 
The difference between $\eqref{quintic}$ with $\eqref{quintic2}$ can be 
compensated by the transformations:\footnote{
Strictly speaking, in the $x_1=1$ patch, we transform as $x_i\rightarrow (a_{i}/a_1)^{-1/5}x_{i}, \ i=2,3,4,5$. 
Transformation for other patches can be obtained quite similarly. 
} 
 \begin{eqnarray}
 x_{i} \rightarrow a_{i}^{-1/5}x_{i} \ (i=1,\cdots, 5) \text{ \ with \ }\psi=-a_{0}(a_{1}a_{2}a_{3}a_{4}a_{5})^{-1/5}. \label{redundant moduli}
 \end{eqnarray}
In this parametrization \eqref{quintic2} one notices an obvious differential relation,
 \begin{eqnarray}
  \le(\fr{\p}{\p a_{0}}\ri)^{5}\frac{\omega_0}{W'}=\prod_{i=1}^{5}\le(\fr{\p}{\p a_{i}}\ri)\frac{\omega_0}{W'}. 
 \end{eqnarray}
We rewrite this equation with \eqref{redundant moduli}, $z=1/\psi^5$ and $\theta=z\p/\p z$:
 \begin{eqnarray}
  \le[\theta^{5}-z\le(5\theta+5\ri)\le(5\theta+4\ri)\le(5\theta+3\ri)\le(5\theta+2\ri)\le(5\theta+1\ri)\ri]\Omega(z)=0 
 \end{eqnarray}
and factorize the differential operator $\theta\mathcal{L}=\theta[\theta^{4}-z\prod_{i=1}^{4}(\theta+i)]\mathcal{L}$.
In case that $\Gamma$ has no boundary,
the period integrals are determined by the four independent solutions of
the Picard-Fuchs equation 
 \begin{eqnarray}
  \le[\theta^{4}-z\prod_{i=1}^{4}\le(\theta+i\ri)\ri]\int_{\Gamma} \Omega(z)=0, \label{P-F_quintic}
 \end{eqnarray} 
because $\mathcal{L}\Omega(z)$ is $\theta$-exact.

In case that $\Gamma$ has boundaries, i.e. $3$-chain, the boundary contributions give rise to the inhomogeneous term of differential equation.\footnote{
The physical meaning of this integral over chain is discussed a little in the next section. 
}

It is not so easy to extend the above algorithm to the integral with
the boundaries,
because the Picard-Fuchs equation is obtained by the fifth-order equation 
rather than the fourth order equation. 
Here we discuss the direct derivation of the fourth order differential 
equation for $\Omega$.
We introduce an integral 
 \begin{eqnarray}
{\mathcal{F}}=\int\frac{\omega_0}{x_{1}x_{2}x_{3}x_{4}x_{5}}\log W'.
\end{eqnarray}
This integral is invariant under the transformation $x_i\rightarrow \lambda x_i, \ (i=1,...,5)$ up to moduli independent term. 
The derivative of $\mathcal{F}$ with respect to the moduli is
well-defined as an integral over $\mathbb{CP}^{4}$. 
It is easy to see 
\begin{eqnarray}
  \le(\fr{\p}{\p a_{0}}\ri)^{5}{\mathcal{F}}=\prod_{i=1}^{5}\le(\fr{\p}{\p a_{i}}\ri){\mathcal{F}}. 
 \end{eqnarray}
Rewriting this equation using (\ref{redundant moduli}),  $\theta=z\p/\p
z$, and a relation, $\Omega =-\theta {\mathcal{F}}$, one can find
the fourth order differential equation: 
 \begin{eqnarray}
  \le[\theta^{4}-z\le(5\theta+4\ri)\le(5\theta+3\ri)\le(5\theta+2\ri)\le(5\theta+1\ri)\ri]\Omega(z)=0.
 \end{eqnarray}
We have seen that a rather trivial relation leads to the Picard-Fuchs 
equation by scaling the integration variables. 
However, in case that $\Gamma$ has boundaries, we cannot
expect that the boundaries do not depend on the moduli. 
Then we should pick up the contributions from the boundaries 
when we consider the rescaling of the variables. 

\subsection{Fourth order differntial equation with inhomogenious term}
The boundaries of the $3$-chain
are specified explicitly after the coordinate
transformations \eqref{redundant moduli}
\cite{MW}. So as to treat the chain integral explicitly,
we have to use the expression of $\Omega$ defined by
\begin{eqnarray}
\Omega &=& -\theta \int \frac{\omega_0}{x_{1}x_{2}x_{3}x_{4}x_{5}}\log W, \nonumber\\
W &=& \frac{1}{5} (x_{1}^5+x_{2}^5+x_{3}^5+x_{4}^5+x_{5}^5)-\psi x_{1}x_{2}x_{3}x_{4}x_{5}.
\end{eqnarray}
In the following, we consider certain scaling method, 
 which is essentially 
equivalent to the above derivation of the Picard-Fuchs equation. 
To perform the rescaling of the coordinates $x_i$, 
we separate $\omega_0$ and $\Omega$ into two parts: 
\begin{eqnarray}
&&\omega_0=\omega_{(1234)}+\omega_{(5)},
\quad \Omega=\Omega_{(1234)}+\Omega_{(5)},
\\
&&\omega_{(1234)}=-x_5dx_1\wedge dx_2\wedge dx_3 \wedge dx_4,
\\
&&\omega_{(5)}=x_4dx_1\wedge dx_2\wedge dx_3 \wedge dx_5
-x_3dx_1\wedge dx_2\wedge dx_4 \wedge dx_5
\nonumber \\
&&\quad\quad\quad
+x_2dx_1\wedge dx_3\wedge dx_4 \wedge dx_5
-x_1dx_2\wedge dx_3\wedge dx_4 \wedge dx_5,
\\
&&\Omega_{(1234)}=\psi\int\frac{\omega_{(1234)}}{W},\quad 
\Omega_{(5)}=\psi\int\frac{\omega_{(5)}}{W}.
\end{eqnarray}
For $\Omega_{(1234)}$, we rescale
$x_i$ $(i=1$, $2$, $3$, $4$) by $\psi\tilde{x}_i$. 
\begin{eqnarray}
\Omega_{(1234)}&=&\psi\int\frac{\omega_{(1234)}}{W}=
-\psi\frac{\partial}{\partial\psi}\int\frac{\omega_{(1234)}}{x_1x_2x_3x_4x_5}
\log W
\nonumber \\
&=&-\psi\frac{\partial}{\partial\psi}
\int\frac{\omega_{(1234)}}{\tilde{x}_1\tilde{x}_2\tilde{x}_3\tilde{x}_4x_5}
\log W
\nonumber \\
&=&-\int\left(5-\frac{x_5^5}{W}\right)\frac{\omega_{(1234)}}{x_1x_2x_3x_4x_5}
+\int\frac{\partial}{\partial x_1}\log W\frac{\omega_{(1234)}}{x_2x_3x_4x_5}
+\int\frac{\partial}{\partial x_2}\log
W\frac{\omega_{(1234)}}{x_1x_3x_4x_5}
\nonumber \\
&&
+\int\frac{\partial}{\partial x_3}\log W\frac{\omega_{(1234)}}{x_1x_2x_4x_5}
+\int\frac{\partial}{\partial x_4}\log W\frac{\omega_{(1234)}}{x_1x_2x_3x_5}. 
\end{eqnarray}
In this computation,
we adopted the following integration formula 
to take into account for the boundary effects properly.
\begin{eqnarray}
&&\psi\frac{\partial}{\partial\psi}\int_a^b \frac{f(x)}{x}dx
=\psi\frac{\partial}{\partial\psi}\int_{a/\psi^m}^{b/\psi^m}
\frac{f(\psi^m\tilde{x})}{\tilde{x}}d\tilde{x}
\nonumber \\
&&
=\int_{a/\psi^m}^{b/\psi^m}\psi\frac{\partial}{\partial\psi}
\frac{f(\psi^m\tilde{x})}{\tilde{x}}d\tilde{x}
-m\int_a^b\frac{\partial}{\partial x}f(x)dx,
\label{boundary_integral_formula}
\end{eqnarray}
where $\tilde{x}:=x/\psi^m$.

For $\Omega_{(5)}$, we rescale only $x_5$ by $\tilde{x}_i/\psi$. 
\begin{eqnarray}
\Omega_{(5)}&=&\psi\int\frac{\omega_{(5)}}{W}
=-\psi\frac{\partial}{\partial\psi}
\int\frac{\omega_{(5)}}{x_1x_2x_3x_4x_5}
\log W
\nonumber \\
&=&\int\frac{x_5^5}{W}\frac{\omega_{(5)}}{x_1x_2x_3x_4x_5}
-\int\frac{\partial}{\partial x_5}\log W
d\log x_1\wedge d\log x_2\wedge d\log x_3\wedge d\log x_5
\nonumber \\
&&+\int\frac{\partial}{\partial x_5}\log W
d\log x_1\wedge d\log x_2\wedge d\log x_4\wedge d\log x_5
\nonumber \\
&&
-\int\frac{\partial}{\partial x_5}\log W
d\log x_1\wedge d\log x_3\wedge d\log x_4\wedge d\log x_5
\nonumber \\
&&
+\int\frac{\partial}{\partial x_5}\log W
d\log x_2\wedge d\log x_3\wedge d\log x_4\wedge d\log x_5,
\end{eqnarray}
where we adopted \eqref{boundary_integral_formula}.
Combining these two contributions, we obtain the expression for
$\Omega$.
\begin{eqnarray}
&&\Omega=\Omega_{(1234)}+\Omega_{(5)}
\nonumber \\
&&=\int\frac{x_5^5}{W}\frac{\omega_0}{x_1x_2x_3x_4x_5}
+5\int\frac{x_5dx_1\wedge dx_2\wedge dx_3\wedge dx_4}{x_1x_2x_3x_4x_5}
\nonumber \\
&&
+\int d\log W\wedge(d\log x_1\wedge d\log x_2\wedge d\log x_3
-d\log x_1\wedge d\log x_2\wedge d\log x_4
\nonumber \\
&&\quad\quad\quad\quad\quad\quad
+d\log x_1\wedge d\log x_3\wedge d\log x_4
-d\log x_2\wedge d\log x_3\wedge d\log x_4
).
\end{eqnarray}

To derive the fourth order differntial equation with repsect to $\psi$, 
we consider the action of $\theta$ on 
$\int\frac{x_5^5}{W}\frac{\omega_0}{x_1x_2x_3x_4x_5}$
term in $\Omega$ first. 
\begin{eqnarray}
&&\theta\int\frac{x_5^5}{W}\frac{\omega_0}{x_1x_2x_3x_4x_5}
\nonumber \\
&&=\int\frac{x_1^5x_5^5}{W^2}\frac{\omega_0}{x_1x_2x_3x_4x_5}
\nonumber \\
&&\quad
+\int d\left(\frac{x_5^5}{W}\right)\wedge
(-d\log x_2\wedge d\log x_3\wedge d\log x_4
+d\log x_2\wedge d\log x_3\wedge d\log x_5
\nonumber \\
&&\quad\quad\quad\quad\quad\quad\quad\quad
-d\log x_2\wedge d\log x_4\wedge d\log x_5
+d\log x_3\wedge d\log x_4\wedge d\log x_5
).
\end{eqnarray}
In the computation, we separated $\omega_0$ and rescaled coordinates
as above. To proceed, we consider the action of $\theta$ on 
$\int\frac{x_1^5x_5^5}{W^2}\frac{\omega_0}{x_1x_2x_3x_4x_5}$
which appears in $\theta^2\Omega$.
\begin{eqnarray}
&&\theta\int\frac{x_1^5x_5^5}{W^2}\frac{\omega_0}{x_1x_2x_3x_4x_5}
\nonumber \\
&&=2\int\frac{x_1^5x_2^5x_5^5}{W^3}\frac{\omega_0}{x_1x_2x_3x_4x_5}
\nonumber \\
&&
\quad
+\int d\left(\frac{x_1^5x_5^5}{W^2}\right)\wedge
(d\log x_1\wedge d\log x_3\wedge d\log x_4
-d\log x_1\wedge d\log x_3\wedge d\log x_5
\nonumber \\
&&\quad\quad\quad\quad\quad\quad\quad\quad
+d\log x_1\wedge d\log x_4\wedge d\log x_5
-d\log x_3\wedge d\log x_4\wedge d\log x_5
).
\end{eqnarray}
In the same manner, we find a term in $\theta^3\Omega$
\begin{eqnarray}
&&\theta\int\frac{x_1^5x_2^5x_5^5}{W^3}\frac{\omega_0}{x_1x_2x_3x_4x_5}
\nonumber \\
&&=3\int\frac{x_1^5x_2^5x_3^5x_5^5}{W^4}\frac{\omega_0}{x_1x_2x_3x_4x_5}
\nonumber \\
&&
\quad
+\int d\left(\frac{x_1^5x_2^5x_5^5}{W^3}\right)\wedge
(-d\log x_1\wedge d\log x_2\wedge d\log x_4
+d\log x_1\wedge d\log x_2\wedge d\log x_5
\nonumber \\
&&\quad\quad\quad\quad\quad\quad\quad\quad
-d\log x_1\wedge d\log x_4\wedge d\log x_5
+d\log x_2\wedge d\log x_4\wedge d\log x_5
),
\end{eqnarray}
and a term in $\theta^4\Omega$
\begin{eqnarray}
&&\theta\int\frac{x_1^5x_2^5x_3^5x_5^5}{W^4}\frac{\omega_0}{x_1x_2x_3x_4x_5}
\nonumber \\
&&=4\int\frac{x_1^5x_2^5x_3^5x_4^5x_5^5}{W^5}\frac{\omega_0}{x_1x_2x_3x_4x_5}
\nonumber \\
&&
\quad
+\int d\left(\frac{x_1^5x_2^5x_3^5x_5^5}{W^3}\right)\wedge
(d\log x_1\wedge d\log x_2\wedge d\log x_3
-d\log x_1\wedge d\log x_2\wedge d\log x_5
\nonumber \\
&&\quad\quad\quad\quad\quad\quad\quad\quad
+d\log x_1\wedge d\log x_3\wedge d\log x_5
-d\log x_2\wedge d\log x_3\wedge d\log x_5
).
\label{fourth_order_quintic}
\end{eqnarray}
The first term in the left hand side of \eqref{fourth_order_quintic}
can be simply given by
\begin{eqnarray}
\int\frac{\omega_0}{W^5}(x_1x_2x_3x_4x_5)^4
=\frac{1}{24}\frac{1}{\psi}\theta\frac{1}{\psi}\theta\frac{1}{\psi}\theta\frac{1}{\psi}\theta\frac{1}{\psi}\theta\Omega.
\end{eqnarray}

From these equations, one finds the fourth order differntial equation 
for $\Omega$.
\begin{eqnarray}
\theta^4\Omega&=&\frac{1}{\psi}\theta\frac{1}{\psi}\theta\frac{1}{\psi}\theta\frac{1}{\psi}\theta\frac{1}{\psi}\theta\Omega
+6\int d\left(\frac{x_1^5x_2^5x_3^5x_5^5}{W^4}\right)\wedge
\frac{\omega_4}{x_1x_2x_3x_5}
+2\theta\int d\left(\frac{x_1^5x_2^5x_5^5}{W^3}\right)\wedge
\frac{\omega_3}{x_1x_2x_4x_5}
\nonumber \\
&&
+\theta^2\int d\left(\frac{x_1^5x_5^5}{W^2}\right)\wedge
\frac{\omega_2}{x_1x_3x_4x_5}
+\theta^3\int d\left(\frac{x_5^5}{W}\right)\wedge
\frac{\omega_1}{x_2x_3x_4x_5}
-\theta^4\int d\log W\wedge\frac{\omega_5}{x_1x_2x_3x_4}.
\nonumber \\
&&
\end{eqnarray}
This equation can be rewritten as follows:
\begin{eqnarray}
&&\theta^4\Omega-\frac{1}{\psi}\theta\frac{1}{\psi}\theta\frac{1}{\psi}\theta\frac{1}{\psi}\theta\frac{1}{\psi}\theta\Omega
=\int d\tilde{\beta},
\\
&&\tilde{\beta}
=6\frac{x_1^4x_2^4x_3^4x_5^4}{W^4}\omega_4
+2\theta\left(\frac{x_1^5x_2^5x_3^5}{W^3}\right)\frac{\omega_3}{x_1x_2x_4x_5}
+\theta^2\left(\frac{x_1^5x_5^5}{W^2}\right)\frac{\omega_2}{x_1x_3x_4x_5}
+\theta^3\left(\frac{x_5^5}{W}\right)\frac{\omega_1}{x_2x_3x_4x_5}
\nonumber \\
&&\quad\quad
-\theta^4\log W\frac{\omega_5}{x_1x_2x_3x_4}
\nonumber \\
&&\quad=
6\frac{x_1^4x_2^4x_3^4x_5^4}{W^4}\omega_4
+6\psi\frac{x_1^5x_2^5x_3x_5^5}{W^4}\omega_3
+6\psi^2\frac{x_1^6x_2^2x_3x_4x_5^5}{W^4}\omega_2
+6\psi^3\frac{x_1^3x_2^2x_3^2x_4^2x_5^7}{W^4}\omega_1
+6\psi^4\frac{x_1^3x_2^3x_3^3x_4^3x_5^4}{W^4}\omega_5
\nonumber \\
&&\quad\quad
+2\psi\frac{x_1^5x_2x_5^5}{W^3}\omega_2
+6\psi^2\frac{x_1^2x_2x_3x_4x_5^6}{W^3}\omega_1
+12\psi^3\frac{x_1^2x_2^2x_3^2x_4^2x_5^3}{W^3}\omega_5
\nonumber \\
&&\quad\quad
+\psi\frac{x_1x_5^5}{W^2}\omega_1
+7\psi^2\frac{x_1x_2x_3x_4x_5^2}{W^2}\omega_5
+\psi\frac{x_5}{W}\omega_5.
\end{eqnarray}
This equation is nothing but the Picard-Fuchs equation 
with the inhomogenious term\footnote{
Precisely speaking, we find 
$\tilde{\beta}$ in \cite{MW} by exchanging $x_1$ and $x_4$
} that was obtained in \cite{MW}.

In this way, we have described an effective alogorithm to obtain the
inohomogeneous term in the Picard-Fuchs equation for $3$-chain integral.
The $3$-chain on mirror quintic is specified by the matrix factorization
\cite{HHP}. The details of the compurtations for $3$-chain integral is
obtained in the celebrated paper \cite{MW}. 
In the next section, we shall propose a method for evalutating
the $3$-chain integral directly via analytic continuation. 
Formally the 3-chain integral from 
the direct computation satisfies 
the inhomogeneous Picard-Fuchs equation 
obtained by this rescaling algorithm.

\subsection{Inhomogeneous Picard-Fuchs equation for double cubic}
It is straightforward to extend our algorithm to the complete intersection models. 
We now discuss the mirror $Y_{3,3}$ of the Calabi-Yau complete intersection $X_{3,3}[1^6]$. 
The ordinal Picard-Fuchs operator and period of this model is discussed in \cite{LibTeit}. 
$Y_{3,3}$ is defined by two homogeneous polynomials of degree three in $\mathbb{CP}^{5}/((\mathbb{Z}_{3})^{2}\times\mathbb{Z}_9)$, 
 \begin{eqnarray}
  &&W_{1}=\fr{1}{3}(x_{1}^{3}+x_{2}^{3}+x_{3}^{3})-\psi x_{4}x_{5}x_{6}=0,\nonumber\\	 
  &&W_{2}=\fr{1}{3}(x_{4}^{3}+x_{5}^{3}+x_{6}^{3})-\psi x_{1}x_{2}x_{3}=0.\label{double_cubic}
 \end{eqnarray}
The holomorphic $3$-form is given by
 \begin{eqnarray}
  \Omega(z)=\mathrm{Res}_{W_{1}=0}\mathrm{Res}_{W_{2}=0}\fr{\omega_{0}}{W_{1}W_{2}},
 \end{eqnarray}
where we have introduced $\omega_{0}$ as the 5-form on the ambient space $\mathbb{CP}^{5}$:
 \begin{eqnarray}
  \omega_{0}=\sum_{i=1}^{6}(-1)^{i}x_{i}dx_{1}\wedge\cdots\wedge\wh{dx}_{i}\wedge\cdots\wedge dx_{6}.
 \end{eqnarray}
Using a small tube $T_{\epsilon}(\Gamma)$ around $\Gamma$ of size $\epsilon$, 
 \begin{eqnarray}
  \int_{\Gamma}\Omega=\int_{T_{\epsilon}(\Gamma)}\fr{\omega_{0}}{W_{1}W_{2}}.
 \end{eqnarray}

We apply our rescaling algorithm to this model.
Representing the defining equations $\eqref{double_cubic}$ as
 \begin{eqnarray}
  &&W'_{1}=a_{1}x_{1}^{3}+a_{2}x_{2}^{3}+a_{3}x_{3}^{3}-a_{7} x_{4}x_{5}x_{6},\nonumber\\
  &&W'_{2}=a_{4}x_{4}^{3}+a_{5}x_{5}^{3}+a_{6}x_{6}^{3}-a_{8} x_{1}x_{2}x_{3},
 \end{eqnarray}
we find the obvious differential relation 
 \begin{eqnarray}
  \prod_{i=1}^{6}\le(\fr{\p}{\p a_{i}}\ri)\Omega=\le(\fr{\p}{\p a_{7}}\ri)^{3}\le(\fr{\p}{\p a_{8}}\ri)^{3}\Omega. \label{eq_cubic}
 \end{eqnarray}
Therefore in this case, we have to factorize two $\theta_{\psi}$'s to obtain the fourth order Picard-Fuchs equation. 
Applying our method to reproduce $\eqref{eq_cubic}$ with $\psi$ derivatives and performing the integration of the exact terms, 
we find the inhomogeneous term of the Picard-Fuchs equation (for the detail of the evaluation, see appendix \ref{details}) is 
 \begin{eqnarray}
  \int_{T_{\epsilon}(\Gamma)}d \beta=
  \int_{T_{\epsilon}(C_{+}-C_{-})}\beta=\fr{4i\pi^{3}}{3\psi^{5}}. \label{111_result}
 \end{eqnarray}
From this and $z:=1/(3\psi)^6$, we can write down the (normalized) inhomogeneous Picard-Fuchs equation as 
 \begin{eqnarray}
  \mathcal{L}_{PF}\mathcal{T}_{B}(z)=\fr{3^{2}}{2^{4}\pi^{2}}z^{1/2}. 
 \end{eqnarray}
The normalization of the domainwall tension is given by 
 \begin{eqnarray}
  \mathcal{T}_{B}(z)=\fr{|GP|}{(2\pi i)^{5}}\psi^{2}\int_{\Gamma}\Omega(z), 
 \end{eqnarray}
where $|GP|$ is the order of the Greene-Plesser orbifold group (so in this case $|GP|=3^{4}$). 
The details of the computations for (the mirror of) $X_{3,3}$ are described 
in Appendix A.

\section{Direct integration via analytic continuation}
In this section, we will discuss the solution of the inhomogeneous Picard-Fuchs equation by a direct computation. 
The approach of the direct computation is studied in \cite{JS2}, 
but we rather discuss the direct computation of the superpotential (or
the tension of BPS domainwall) itself by performing the analytic continuation of the period integral. 
The advantage of our method is that it reproduces the whole expression of the superpotential difference, 
whereas the Picard-Fuchs equation only determines it up to the periods. 

\subsection{Mirror of quintic $X_5[1^5]$}
The defining equation of the mirror quintic Calabi-Yau $3$-fold $Y_5$ is given in (\ref{quintic}). 
Let $C_\pm$ be 
\begin{equation}
C_{\pm}=\{x_1+x_2=0,\quad x_3+x_4=0,\quad x_5^2\pm \sqrt{5\psi} x_1x_3=0,\}, \label{curve_quintic}
\end{equation}
then they are B-brane mirror to the real Lagrangian submanifold in
A-side, which is defined by the fixed locus of the antiholomorphic involution. 
This B-brane can be obtained by use of the matrix factorization method and the grade restriction rule \cite{MW,HHP}. 
Moreover $C_+$ and $C_-$ are homologous each other and there exists a $3$-chain $\Gamma$ which interpolates between these curves $C_\pm$. 
Physically, $C_\pm$ correspond to two supersymmetric vacua of an $\mathcal{N}=1$ supersymmetric theory on the $D5$-brane worldvolume\footnote{
This $D5$-brane locates entire non-compact $\mathbb{R}^{1,3}$ and wraps the curves $C_\pm$ in Calabi-Yau $3$-fold. 
} 
and we can find the BPS domainwall which wraps $\Gamma$, with boundaries on $C_\pm$. 
The tension of a BPS domainwall between the two vacua is equal to the difference of superpotentials of $C_\pm$, 
and given by the integral of holomorphic $3$-form over $3$-chain $\Gamma$ \cite{Witten}. 
In a mathematical terminology, it determines a Griffiths' normal function of the variation 
of the mixed Hodge structure (see e.g. \cite{Griffiths,Griffiths2,Green} and the references therein) 
and the superpotentials have information about the obstruction of curves $C_\pm$ \cite{KKLM}.

We shall perform the integration of the holomorphic $3$-form
in the patch of $x_1=1$. The computation of the another patch $x_3=1$ 
gives the same result under the exchange of the coordinates 
$x_1\leftrightarrow x_3$ and $x_2\leftrightarrow x_4$. 
On the $x_1=1$ patch, the period integrals yield to 
\bea\label{3chain-integral}
\Pi = \frac{5^3 \psi}{(2\pi i)^4}
\int_{T_{\epsilon} (\Gamma)} \frac{dx_{2}dx_{3}dx_{4}dx_{5}}{W\big|_{x_{1}=1}}.
\ena
The fundamental period $\Pi_0$ is obtained by integrating over the tubular domain of the fundamental cycle, 
$T_{\epsilon}(\Gamma_0)=\gamma_2\times\gamma_3\times\gamma_4\times \gamma_5$, 
where $\gamma_{\ell}$ encircle the complex coordinates $x_{\ell}$ \cite{CDGP}. 
Because of the difficulty of finding $3$-cycles whose periods include doubly logarithmic terms $(\log z)^2$, 
one usually calculates the periods by using the Picard-Fuchs differential equations which govern several periods \cite{Mo}. 
On the other hand, the domainwall tension is given by the integration 
over the tubular domain of the $3$-chain $\Gamma$ whose boundary is $\partial \Gamma=C_+-C_-$. 
It will be desirable to compute the $3$-chain integral directly. 
because it is expected from the A-model that the superpotential difference contains at most single logarithm \cite{Wa1},

Taking into account for the resolution of the singularities \cite{MW}, 
we introduce the good coordinates
\bea
T&=&x_1^{-1}x_{2}, \ \ \ 
X=x_{1}x_{3}^{-2}x_{4}^3x_{5}^{-2},\nonumber\\
Y&=&x_{1}^{-5}x_{5}^5, \ \ \ 
Z=x_{1}x_{3}^3x_{4}^{-2}x_{5}^{-2}.\label{local coordinate quintic}
\ena
In these local coordinates, the defining equation in the $x_{1}=1$ patch becomes
\bea
W=\frac{1}{5}[1+T^5+(X^2Z^3+X^3Z^2)Y^2+Y(1-5\psi TXZ)], \label{eq:definingeq}
\ena
and the brane loci
$C_{\pm}$ correspond to  $T=-1$, $X=-Z=\pm \frac{1}{\sqrt{5\psi}}$. 

For the evaluation of the $3$-chain integrals, 
we further introduce the ``polar coordinates'' for $X$ and $Z$ as 
\begin{eqnarray}
X=\frac{\zeta w}{(5\psi T)^{1/2}}, \quad 
Z=-\frac{\zeta^{-1}w}{(5\psi T)^{1/2}}, 
\end{eqnarray}
where the coordinate $w$ covers the whole complex plane whereas 
the coordinate $\zeta$ covers half of the plane.
In these coordinates, the defining equation can be rewritten as
 \bea
 W=\frac{1}{5}[1-T^5+(5\psi)^{-5/2}(\zeta-\zeta^{-1})w^5T^{-5/2}Y^2+Y(1-w^2)],
 \ena
and $C_{\pm}$ correspond to 
\begin{equation}
\zeta=1, \ w=\pm 1, \ T=-1.\label{brane quintic}
\end{equation}

The period integral \eqref{3chain-integral} 
can be expressed in terms of these local coordinates as
\begin{eqnarray}
\Pi=
\frac{10}{(2\pi i)^4}\int \frac{d\zeta}{\zeta}\frac{dT}{T}wdwdY
\frac{1}{1-T^5+(5\psi)^{-5/2}(\zeta-\zeta^{-1})w^5T^{-5/2}Y^2+Y(1-w^2)}. 
\end{eqnarray}
Integration over $Y$ can be easily performed (we pick up one of two simple poles) and we find 
\bea
\Pi =\frac{10}{(2\pi i)^3}\int\frac{d\zeta}{\zeta} \frac{dT}{T}wdw\frac{1}{\sqrt{(1-w^2)^2
-4w^5(\zeta-\zeta^{-1})(T^{5/2}-T^{-5/2})z^{1/2}}},
\ena
where we have defined $z=1/(5\psi)^5$.

In the large moduli limit, the $3$-chain integrals can be expanded as
\bea
\Pi =&{}&10\sum_{n=0}^{\infty}
\frac{(2n)!}{(n!)^2}\int\frac{dw}{2\pi i}\frac{w^{5n+1}}{(1-w^2)^{2n+1}}\nonumber\\
&{}&\times\int\frac{d\zeta}{(2\pi i)\zeta}
(\zeta-\zeta^{-1})^n\int\frac{dT}{(2\pi i)T}(T^{5/2}-T^{-5/2})^n z^{n/2}.
\ena
This expression is very useful, because all of the integrals are separated. 
This separation enables us to determine paths of the integration independently. 
This formula is common for cycle/chain integrals and we choose the appropriate contours of each variable according to what we want to calculate. 
Basically, the difference between the fundamental period and the chain integral appears as a choice of paths of the $w$-integration. 

\begin{figure}
\begin{center}
\includegraphics[width=5 cm,clip]{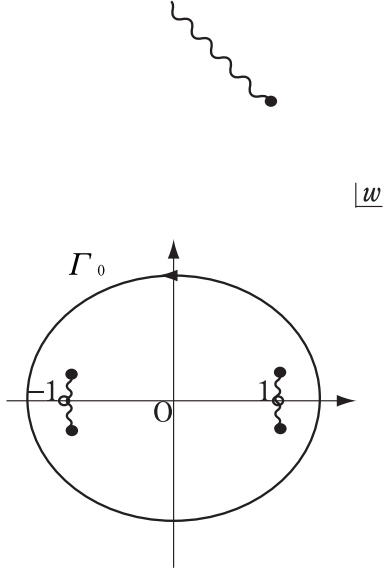}
\end{center}
\caption{Singular points and the contour representing the fundamental cycle}
\label{fig:zplane1}
\end{figure}

There are six singular points in the $w$-integration. 
In the small $z$ limit (i.e. the large moduli limit in A-model), 
four of them are located near $w=\pm 1$ and one of them is located near infinity 
and the last one is at infinity.  
The fundamental cycle encircles around the infinity in the large moduli limit.  
Therefore, we identify the contour for fundamental cycle $\Gamma_0$ 
which encircles around $w=-1$ and $w=1$ drawn in Fig.\ref{fig:zplane1}. 

As a contour for the $\zeta$-coordinate, we choose 
$\zeta=e^{i\theta}$, $-\pi/2<\theta<\pi/2$. For $T$ we choose 
$T=e^{i\phi}$, $0<\phi<2\pi/5$, because of the Greene-Plessor orbifold group action. 
Non-trivial $\Bbb{Z}_5$ actions on $(T,X,Y,Z)$ coordinates are expressed as the following charges: 
\begin{eqnarray}
g_1=(3,1,0,1),\quad g_2 =(4,3,0,3), 
\end{eqnarray}
and the $Y$ coordinate is a singlet. 

To check the validity of the above contour, 
we are going to evaluate the fundamental period. 
The integration over $\zeta$ vanishes unless $n=2m$ $(m=0,1,\cdots)$, and we find 
\bea
\int\frac{d\zeta}{2\pi i \zeta}(\zeta-\zeta^{-1})^{2m}=(-1)^m\frac{1}{2}\frac{(2m)!}{(m!)^2}. 
\ena
The integration over $T$ can be performed and we find 
\bea
\int\frac{dT}{2\pi i T}(T^{5/2}-T^{-5/2})^{2m}=(-1)^m\frac{1}{5}\frac{(2m)!}{(m!)^2}. 
\ena
We can also perform the integral over $w$  by changing $w=1/x$ and evaluating the pole at $x=0$. 
\bea
\int_{\Gamma_0}\frac{dw}{2\pi i}\frac{w^{10m+1}}{(1-w^2)^{4m+1}}=\frac{(5m)!}{(4m)!m!}. 
\ena
The contour $\Gamma_0$ is shown in Fig.\ref{fig:zplane1}. 
Collecting these results, we obtain the following well-known result \cite{CDGP}: 
\bea
\Pi_0=\sum_{m=0}^\infty \frac{(5m)!}{(m!)^5}z^m. 
\ena
This fact supports the validity of our contour for each variables. 

\begin{figure}
\begin{center}
\includegraphics[width=5 cm,clip]{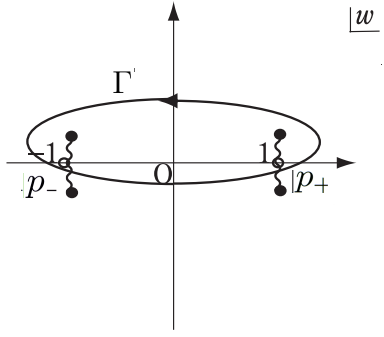}
\end{center}
\caption{The contour connecting two boundaries}
\label{fig:zplane}
\end{figure}

We are now going to evaluate the $3$-chain integral with boundaries $\partial \Gamma=C_+-C_-$.
Because the brane is located at \eqref{brane quintic}, 
we must be careful when considering the $w$-contour. 
Before resolving the Hirzebruch-Jung singularity of $\{x_1+x_2=0=x_3+x_4\}/\mathbb{Z}_5$, one finds two intersection points of 
$C_+\cap C_-$. 
One of them is located in the $x_1=1$ patch, we denote it by $p$. 
After the resolution the Hirzebruch-Jung singularity \cite{MW}, 
each singular point is replaced by the rational curves and we denote 
the intersections of $C_\pm$ with the such curves by $p_\pm$. 
In the $w$-plane they correspond to $w=\pm 1$. 
We can easily expect that the contour, which corresponds to the chain connecting $p_+$ and $p_-$, can be written as in Fig.\ref{fig:zplane}. 
The contour is bounded by the two boundaries because it cannot be removed from these singular points. 
Since this contour integral encircles cut, it can be represented as two times the line integral connecting two boundaries. 

Now we must be careful about the covering of the coordinates.  The patch $x_1=1$ cannot cover the entire boundary. 
Therefore, we should add another contribution to the boundary which can be obtained in the patch $x_3=1$. 
It is enough to cover the entire boundary by these two patches. 
However, the form of the other local coordinates is identical to the one used here. 
Eventually, we claim that the line integral representing the $3$-chain integral connecting two boundaries is 
four times (i.e. two patches and two line integrals) the line integral connecting $w=-1$ and $w=1$ in the large moduli expansion.

\begin{figure}
\begin{center}
\includegraphics[width=5 cm,clip]{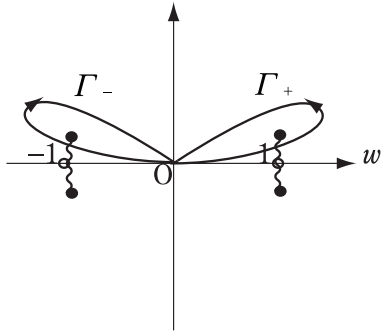}
\end{center}
\caption{The two contours $\Gamma_\pm$}
\label{fig:zplane2}
\end{figure}

Then, in order to evaluate the $w$-integration, we consider the analytic continuation of the above period integral.
We replace the discrete sum with respect to $n$ by the contour integral with respect to $s$.
This is achieved by the Barnes integral formula, so the $3$-chain integral becomes 
\bea
\Pi &=&40\int\frac{ds}{2\pi i}\frac{\pi \cos(\pi s)}{\sin(\pi s)}\frac{\Gamma(2s+1)}{\Gamma(s+1)^2}\nonumber\\
&\times&\int_{-1}^{1}\frac{dw}{2\pi i}\frac{w^{5s+1}}{(1-w^2)^{2s+1}}\int\frac{d\zeta}{2\pi i\zeta}(\zeta-\zeta^{-1})^s\int\frac{dT}{2\pi iT}(T^{5/2}-T^{-5/2})^s z^{\frac{s}{2}},
\ena
where the contour with respect to $s$ encircles non-negative integers. 
Next we {\it assume} that 
we can separate $\Gamma$ into two parts $\Gamma_+$ and $\Gamma_-$, 
which are intuitively considered as the contributions of $C_+$ and $C_-$ respectively 
(shown in Fig.\ref{fig:zplane2}). 
We denote the $\Gamma_+$ contribution by $\Pi_+$ and $\Gamma_-$ contribution by $\Pi_-$. 
Since the boundary $C_+$ corresponds to $T=1$, $\zeta=1$ and $w=+1$, 
we will identify the chain integral $\Pi_+$ as two times the line integral starting from $w=0$ to $w=1$. 
Similarly, the integral $\Pi_-$ can be written as the line integral from $-1$ to $0$ 
and we can obtain this $\Pi_-(z)$ by the relation $\Pi_-(z^{1/2})=\Pi_+({-z^{1/2}})$. 
Namely, we claim 
\bea
\Pi=\Pi_--\Pi_+=\int_{\Gamma_-}-\int_{\Gamma_+}=2\int_{-1}^0 dw(\cdots)-2\int_0^1 dw(\cdots). 
\ena
This is the assumption we should make. 

Later we will adopt the similar prescription for the complete intersection Calabi-Yau case (double cubic model), 
in which case the situation is a little bit different. 
Nevertheless we will see that this prescription reproduces the known results for the chain integral. 

So the integral we should evaluate is 
\bea
\Pi_+ &=&40\int\frac{ds}{2\pi i}\frac{\pi \cos(\pi s)}{\sin(\pi s)}\frac{\Gamma(2s+1)}{\Gamma(s+1)^2}\nonumber\\
&{}&\quad\times\int_0^1\frac{dw}{2\pi i}\frac{w^{5s+1}}{(1-w^2)^{2s+1}}\int\frac{d\zeta}{2\pi i\zeta}(\zeta-\zeta^{-1})^s\int\frac{dT}{2\pi iT}(T^{5/2}-T^{-5/2})^s z^{\frac{s}{2}}.
\label{quinticintegral}
\ena
We are now going to evaluate each integral contained in the above formula. 
The integral over $w$ can be evaluated as 
\bea
\int_0^1 dw \frac{w^{5s+1}}{(1-w^2)^{2s+1}}=\frac{1}{2}\frac{\pi}{\sin (2\pi s)}\frac{\Gamma({\frac{5s}{2}+1})}{\Gamma(2s+1)\Gamma(\frac{s}{2}+1)}.
\ena
$\zeta$ has a parametrization of the half circle $\zeta=e^{i\theta}$, $-\pi/2<\theta<\pi/2$ 
and the $\zeta$ integral can be evaluated 
as half the value of the integral of the contour. The result is 
\bea
\frac{1}{2\pi i}\int\frac{d\zeta}{\zeta}(\zeta-\zeta^{-1})^s=\frac{1}{2}\cos\left(\frac{\pi s}{2}\right)\frac{\Gamma(s+1)}{\Gamma(\frac{s}{2}+1)^2}. \label{zeta_int}
\ena
As noted before, because of the orbifold structure of the variable $T$, the integration region of $T$ is $T=e^{i\phi}$, $0<\phi<2\pi/5$. 
Therefore, the integral of $T$ leads to 
\bea
\frac{1}{2\pi i}\int\frac{dT}{T}(T^{5/2}-T^{-5/2})^s=\frac{e^{\pi is/2}}{5}\frac{\Gamma(s+1)}{\Gamma(\frac{s}{2}+1)^2}. \label{T_int}
\ena
It is now easy to perform the integrations to find
\bea
\Pi_+=\frac{1}{2\pi i}\int\frac{ds}{2\pi i}\frac{\pi \cos(\pi s)}{\sin(\pi s)}
e^{\pi i s/2}\frac{\Gamma(\frac{5}{2}s+1)}{\Gamma(\frac{s}{2}+1)^5}\left(\frac{2\pi \cos(\frac{\pi s}{2})}{\sin(2\pi s)}\right)z^{s/2}.
\ena
The integral of $s$ has double poles for $s=2n$ and single poles for $s=2n+1$ $(n=0, 1, 2, \cdots)$. By evaluating these poles, we finally have
\bea
\Pi_+=\frac{1}{4\pi i}\varpi_1 + \frac{1}{4}\varpi_0 + \frac{1}{4}\sum_{n=0}^\infty \frac{\Gamma(n+\frac{7}{2})}{\Gamma(n+\frac{3}{2})^5}z^{n+1/2}. 
\ena
Here $\varpi_0$ is the fundamental period and $\varpi_1$ is the logarithmic period given by 
\bea
\varpi_1(z)={\cal\varpi}_0(z)\log z +\sum_{n=0}^\infty\frac{\Gamma(5n+1)}{\Gamma(n+1)^5}(5\Psi(5n+1)-5\Psi(n+1))z^n,
\ena
where $\Psi$ denotes the digamma function. 
By using the relation $\Pi_-(z^{1/2})=\Pi_+({-z^{1/2}})$, we get 
\bea
\Pi_-=\frac{1}{4\pi i}\varpi_1 -\frac{1}{4}\varpi_0 - \frac{1}{4}\sum_{n=0}^\infty \frac{\Gamma(n+\frac{7}{2})}{\Gamma(n+\frac{3}{2})^5}z^{n+1/2}.
\ena
This result agrees with the form given in \cite{MW}. 
This agreement can be used as the justification of the choice of the line integral we have made for $w$-integral. 

By the results of  \cite{Wa1,PaSoWa,MW}, under the mirror map 
\bea
\log q(z)=2\pi i t(z)=\frac{\varpi_1(z)}{\varpi_0(z)},
\ena
$\Pi_\pm(z)$ 
are identical to ``the A-model normal function'' \cite[\S 3.2]{MW} of the real quintic 
after suitable normalization of the holomorphic $3$-form: 
\bea
\frac{\Pi_{\pm}(z(q))}{\varpi_0(z(q))}=
\frac{t}{2}\pm\left(\frac{1}{4}+\frac{1}{2\pi^2}\sum_{k,d\; {\rm odd}}
\frac{2n_d^{(0,\mathrm{real})}}{k^2}q^{kd/2}\right), 
\ena
where $n_d^{(0,\mathrm{real})}$ 
denote real BPS numbers \cite{OV,Wa3} (which are the half of disk instanton numbers). 
This number should be integer by the enumerative interpretation and we can exactly confirm this property \cite{Wa1,PaSoWa,MW}. 

\subsection{Mirror of double cubic $X_{3,3}[1^6]$}
Now we compute the period for the mirror geometry of the double cubic $X_{3,3}[1^6]$.
As is the case with mirror quintic, the computation of fundamental period works well, so we will concentrate on the chain integral only. 
The mirror Calabi-Yau $Y_{3,3}$ is the complete intersection 
in $\mathbb{CP}^5/((\mathbb{Z}_{3})^2\times \mathbb{Z}_{9})$ and 
the defining equations are given by
\begin{eqnarray}
W_{1}&=&\frac{1}{3}(x_{1}^3+x_{2}^3+x_{3}^3)-\psi x_{4}x_{5}x_{6}, \nonum\\
W_{2}&=&\frac{1}{3}(x_{4}^3+x_{5}^3+x_{6}^3)-\psi x_{1}x_{2}x_{3}.\label{double cubic def}
\end{eqnarray}
The D5-brane is wrapping around the curves $C_{\pm}$
which are given by the intersection with hyperplanes $x_1+x_2=0$ and $x_3+x_4=0$,
\begin{eqnarray}
C_\pm=\{x_1+x_2=0,\quad x_4+x_5=0,\quad x_3^3-3\psi x_4x_5x_6=0,\quad x_6^3-3\psi x_1x_2x_3=0\}. \label{double cubic curve def}
\end{eqnarray}
Since the orbifold group is $(\mathbb{Z}_3)^2\times\mathbb{Z}_{9}$, the $3$-chain integral in the $x_5=1$ patch is 
\begin{equation}
\Pi=\frac{3^4}{(2\pi i)^5}\int \frac{dx_1dx_2dx_3dx_4dx_6}{W_1W_2}. 
\end{equation}
To describe the curves we should introduce the resolved coordinates 
to which the Greene-Presser orbifold group acts nicely. 
In the patch $x_5=1$, the coordinates are 
\begin{eqnarray}
x_1^3=XZ^2Y^4,\quad x_2^3=X^2Z Y^4,\quad x_3^3=Y,\quad x_4=T,\quad
x_5=1,\quad x_6=UY. 
\label{double_cubic_1}
\end{eqnarray}
The coordinate for another patch $(x_2=1)$ is also found by exchanging 
$(x_1,x_2,x_3)$ $\leftrightarrow$ $(x_4,x_5,x_6)$, 
and the curve $C_{\pm}$ are completely covered by these two patches. 

In these local coordinates, the location of the brane $C_\pm$ is specified by 
\begin{eqnarray}
T=-1,\quad U=-\frac{1}{3\psi},\quad X=-Z=\pm\frac{1}{(3\psi)^2}. 
\end{eqnarray}
Now as in the case of quintic, we consider $\Pi_+$, the contribution from $C_+$. 
The $3$-chain integral $\Pi_+$ can be expressed as 
\begin{eqnarray}
\Pi_+=
\frac{3^3\psi^2}{(2\pi i)^5}
\int\frac{dXdYdZdTdU}
{\left[1-3\psi TU+(XZ^2+X^2Z)Y\right]\left[(U^3-3\psi XZ)Y+1+T^3\right]}.
\end{eqnarray}
One can perform the integration with respect to the variable $Y$ in the
$3$-chain integral by picking up the residue. 
\begin{eqnarray}
\Pi_+=\frac{3^3\psi^2}{(2\pi i)^4}
\int \frac{dXdZdTdU}
{(U^3-3\psi XZ)(1-3\psi TU)-(XZ^2+X^2Z)(1+T^3)}.
\end{eqnarray}
To perform the contour integrations for the remaining variables, we introduce the polar coordinates: 
\begin{eqnarray}
&&X=\frac{w \zeta v^{3/2}}{(3\psi)^{2}t^{3/2}},\quad 
Z=-\frac{w \zeta^{-1} v^{3/2}}{(3\psi)^{2}t^{3/2}},
\nonumber \\
&&U=-\frac{v}{3\psi t}, \quad
T=-t.\label{double_cubic_2}
\end{eqnarray}
Then the $3$-chain integral can be expressed as
\begin{eqnarray}
&&\Pi_{+}=\frac{6}{(2\pi i)^5}
\int\frac{w dw  dv dt d\zeta}{t\zeta}
\frac{1}{(1-w^2)(v-1)
-w^3v^{3/2}(\zeta-\zeta^{-1})(t^{3/2}-t^{-3/2})(3\psi)^{-3}}.
\nonumber \\
&&
\end{eqnarray}
Near the large complex structure limit point, this expression can be expanded with respect to $z=1/(3\psi)^6$, and
rewritten as the Barnes integral form
\begin{eqnarray}
\Pi_{+}
&=&
\frac{6}{2\pi i}\int\frac{ds}{2\pi i}\frac{\pi \cos(\pi s)}{\sin(\pi s)}z^{s/2}
\int \frac{dw}{2\pi i}
\frac{w^{3s-1}}{(1-w^2)^{s+1}}
\int \frac{dv}{2\pi i}
\frac{v^{3s/2}}{(v-1)^{s+1}}
\nonumber \\
&&\times \int\frac{d\zeta}{2\pi i\zeta}(\zeta-\zeta^{-1})^s
\int\frac{dt}{2\pi it}(t^{3/2}-t^{-3/2})^s.
\label{barnesform}
\end{eqnarray}
Contrary to the quintic case where the cut structure is present for $w$, there is no cut structure for the integral here. 
However we apply the same prescription as quintic case and identify the
integral over $w$ as a line integral. 
Thus, we can find the contribution from $\Pi_+$ in the $x_5=1$ patch as a line integral from $w=0$ to $w=1$. 
Since we also have the contribution from the patch $x_2=1$ which can be converted to the same form as (\ref{barnesform}), 
we identify $\Pi_+$ as two times the line integral from $w=0$ to $w=1$. 

\begin{figure}
\begin{center}
\includegraphics[width=5 cm,clip]{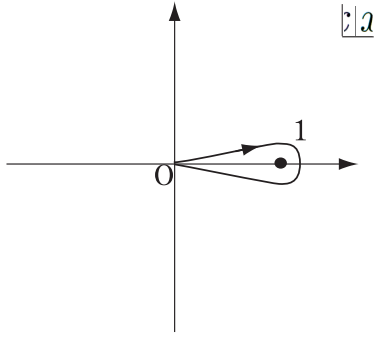}
\end{center}
\caption{The contour of $x$}
\label{fig:tplane1}
\end{figure}

For the several integrations, we shall frequently use the following formula, 
\bea
\oint \frac{dx}{2\pi i}\frac{x^{m s/2}}{(1-x)^{Ns+1}}&=&\frac{1-e^{2N\pi is}}{2\pi i}\int_{0}^{1}dx \frac{x^{m s/2}}{(1-x)^{Ns+1}}\nonumber\\
&=&e^{N\pi is}\frac{\Gamma(\frac{ms}{2}+1)}{\Gamma((\frac{m}{2}-N)s+1)\Gamma(Ns+1)},
\ena
where the contour is chosen as in Fig.\ref{fig:tplane1}.

Taking into account for the Greene-Presser group action $G\simeq
(\mathbb{Z}_{3})^2\times \mathbb{Z}_9$, we specify the integral contours
for $(\zeta,t)$. Each variables are parametrized as 
$(\zeta,t)=(e^{i\theta},e^{i\phi})$
where $-\pi/2\leq \theta\leq \pi/2$ and $0\leq \phi\leq 2\pi/3$.
Therefore the values of integrals $\zeta$ and $t$ are just half and $1/3$ of the result of contour in Fig.\ref{fig:tplane1}, respectively. 
Then each integral in the $3$-chain integral are evaluated as follows:
\begin{eqnarray}
&& \int_0^1
 \frac{dw}{w}
\frac{w^{3s}}{(1-w^2)^{s+1}}=\frac{\pi}{2\sin(\pi s)}
\frac{\Gamma(\frac{3s}{2}+1)}{\Gamma(s+1)\Gamma(\frac{s}{2}+1)},
\nonumber \\
&&\int \frac{dv}{2\pi i}\;\frac{v^{3s/2}}{(v-1)^{s+1}}
=\frac{\Gamma(\frac{3s}{2}+1)}{\Gamma(\frac{s}{2}+1)\Gamma(s+1)},
\nonumber \\
&&\int\frac{d\zeta}{2\pi i\zeta}(\zeta-\zeta^{-1})^s=\frac{\cos(\frac{\pi s}{2})}{2}\frac{\Gamma(s+1)}{\Gamma(\frac{s}{2}+1)^2},
\nonumber \\
&&\int\frac{dt}{2\pi i} t^{-\frac{3}{2}s-1}(t^3-1)^s
=\frac{e^{\pi is/2}}{3}
\frac{\Gamma(s+1)}{\Gamma(\frac{s}{2}+1)^2}.\label{double_cubic_integrals}
\end{eqnarray}
Collecting all contributions, one finds
\begin{eqnarray}
\Pi_+=\frac{1}{2\pi i}\int\frac{ds}{2\pi i}
\frac{\pi\cos(\pi s)}{\sin(\pi s)}e^{\pi is/2}\frac{\pi\cos(\frac{\pi s}{2})}{\sin(\pi s)}
\frac{\Gamma(\frac{3s}{2}+1)^2}{\Gamma(\frac{s}{2}+1)^6}z^{s/2}. 
\end{eqnarray}
There are simple poles at $s=2n+1$ and double poles at $s=2n$.
After the reside integrals on $s$, we obtain
\begin{eqnarray}
\Pi_+=\frac{1}{4\pi i}\varpi_1+\frac{1}{4}\varpi_0+\frac{1}{4}\tau, 
\end{eqnarray}
where the fundamental period $\varpi_0$, the logarithmic period
$\varpi_1$, and the remaining term $\tau$ are given as follows: 
\begin{eqnarray}
&&\varpi_0=\sum_{n=0}^{\infty}\frac{\Gamma(3n+1)^2}{\Gamma(n+1)^6}z^n, \\
&&\varpi_1=\varpi_0\log z+6\sum_{n=0}^{\infty}\frac{\Gamma(3n+1)^2}{\Gamma(n+1)^6}z^n[\Psi(3n+1)-\Psi(n+1)], \\
&&\tau=\sum_{n=0}^{\infty}\frac{\Gamma(3n+\frac{5}{2})^2}{\Gamma(n+\frac{3}{2})^6}z^{n+1/2}. 
\end{eqnarray}
$\Pi_-$ can be also obtained in the same way as the quintic case.

\subsection{Mirror of $X_{12}[1^2,2^2,6]$}
In this subsection we discuss a rather non-trivial example, two-parameter Calabi-Yau hypersurface, $X_{12}[1^2,2^2,6]$.\footnote{
Several analyses of (closed) mirror symmetry for $2$-moduli Calabi-Yau are in \cite{KS,HKTY,HKTY2,COFKM1,CFKM2,BKK}. 
} 
This model is expected to have the $\mathbb{Z}_3$-structure of the open
string vacuum, and $\mathbb{Z}_3$-instanton numbers are calculated in \cite{Walcher} by using open mirror symmetry.\footnote{
The similar structure ($\mathbb{Z}_k$-vacua ($k\neq 2$)) is also observed in $X_{4,4}[1^4,2^2]$ and $X_{6,6}[1^2,2^2,3^2]$. 
Although the A-model picture is so far missing, there are the enumerative predictions for real BPS numbers \cite{Walcher}. 
}
The solution of the inhomogeneous Picard-Fuchs equation is complicated,
but we will see that it can be obtained  a rather simply by our method. 
We also note that (probably a different sector of) this model is discussed in \cite{JS2} by another technique, simple direct integration. 

The defining polynomial of mirror $Y_{12}$ is given by
\bea
W=\frac{1}{12}x_{1}^{12}+\frac{1}{12}x_{2}^{12}-\frac{1}{6}x_{3}^{6}-\frac{1}{6}x_{4}^{6}+\frac{1}{2}x_{5}^{2}-\psi x_{1}x_{2}x_{3}x_{4}x_{5}-\frac{\phi}{6}(x_{1}x_{2})^{6}, 
\ena
and the Greene-Plessor orbifold group is $(\mathbb{Z}_6)^2\times\mathbb{Z}_2$. 
Note that the signs for the third and forth monomials are changed by phase transformations for convenience. 
Following \cite{Walcher}, we choose the boundary as the intersection with hyperplanes such as 
\bea
x_{1}^2=2^{1/6}x_{3},\qquad x_{2}^2=2^{1/6}x_{4}.
\ena
Then the boundaries can be specified by 
\bea
x_{5}=\alpha_\pm(x_{1}x_{2})^3, 
\ena
where $\alpha_\pm$ are the two solutions of 
\bea
\frac{1}{2}\alpha^2-2^{-1/3}\psi\alpha -\frac{\phi}{6}=0. 
\ena
We choose the local coordinate as 
\bea
x_{1}=1, \ x_{3}=2^{-1/6}T, \ x_{5}^2=Y, \ x_{2}^6=YX^4, \ x_{4}^6=\frac{1}{2}X^2Z^6,
\ena
then the defining polynomial can be written as follows: 
\bea
2W=\frac{1}{6}(1-T^6)+\frac{1}{6}Y^2(X^8-X^2Z^6)+Y\left(1-2^{2/3}\psi XZT-\frac{1}{3}X^4\right).
\ena
By introducing the following polar coordinates 
\bea
X=\frac{w^{1/2}\zeta^{-1/12}}{(2^{2/3}\psi T)^{1/2}}, \ Z=\frac{w^{1/2}\zeta^{1/12}}{(2^{2/3}\psi T)^{1/2}}, \ T^2=t,
\ena
the defining equation can be rewritten as 
\bea
2W=\frac{1}{6}(1-t^3)+\frac{1}{6}Y^2\frac{1}{(2^{2/3}\psi)^4}\frac{w^4}{t^2\zeta^{2/3}}(1-\zeta)+Y\left(1-w-\frac{\phi}{3(2^{2/3}\psi)^2}\frac{w^2}{t\zeta^{1/3}}\right).
\ena
In these variables, the location of the boundaries are specified by
\bea
t=\eta \ (\eta^3=1),\quad \zeta=1,\quad w=1. 
\ena
The period integral becomes
\bea
\Pi=\frac{6^2\times2}{(2\pi i)^4}\int \frac{dx_{2}dx_{3}dx_{4}dx_{5}}{W}
=\frac{1}{(2\pi i)^4}\int\frac{dt}{t}\int\frac{d\zeta}{\zeta}\int dw\int dY\frac{1}{2W}.
\ena
By making integral over $Y$, we have 
\bea
\Pi&=&\frac{1}{(2\pi i)^3}\int\frac{dt}{t}\int\frac{d\zeta}{\zeta}\int dw\sum_{n=0}^\infty\frac{\Gamma(2n+1)}{\Gamma(n+1)^2}
\nonumber\\
&\times&\frac{(1-t^3)^nt^{-2n}(1-\zeta)^n\zeta^{-2n/3}w^{4n}}{\left(1-w-\frac{\phi}{3(2^{2/3}\psi)^2}\frac{w^2}{t\zeta^{1/3}}\right)^{2n+1}}\left(\frac{1}{12^{2}2^{8/3}\psi^4}\right)^n.
\ena
Expanding with respect to the variable $\phi$ and 
considering the analytic continuation of the summation with respect to $k+2n$ to the integration with respect to $s$, we have 
\bea
\Pi&=&\frac{1}{(2\pi i)^3}\int\frac{ds}{2\pi i}\frac{\pi \cos(\pi s)}{\sin(\pi s)}\sum_{n=0}^\infty\frac{\Gamma(s+1)}{\Gamma(s-2n+1)\Gamma(n+1)^2}\nonumber\\
&\times&\int dt\frac{(1-t^3)^n}{t^{s+1}}\int d\zeta \frac{(1-\zeta)^n}{\zeta^{s/3+1}}\int dw\frac{w^{2s}}{(1-w)^{s+1}}
\left(\frac{\phi}{3\cdot2^{4/3}\psi^2}\right)^s\left(\frac{3^2}{12^2\phi^2}\right)^{n}.
\ena
We perform the integration over $t$ as line integral from $0$ to $\eta^{-1}$ $(\eta^3=1)$. For $\zeta$, we choose the integral from $0$ to $1$. 
The integration of $w$ is the contour integral around $w=1$. 
In this way, we obtain 
\bea
\Pi&=&3\sum_{n=0}^\infty\int\frac{ds}{2\pi i}\frac{\pi \cos(\pi s)}{\sin(\pi s)}\left(\frac{\frac{\pi}{3}}{\sin(\frac{\pi s}{3})}\right)^2\nonumber\\
&\times&\frac{\Gamma(2s+1)}{\Gamma(\frac{s}{3}+1)^2\Gamma(n-\frac{s}{3}+1)^2\Gamma(s-2n+1)\Gamma(s+1)}\eta^{s}y^{s/3}z_2^n,
\ena
where $y=\frac{\phi}{3\cdot 2^{4/3}\psi^2}$ and $z_2=\frac{3^2}{12^2\phi^2}$. 

By evaluating poles of the $s$-integration, we get a rather complicated result obtained in \cite{Walcher}. 
It is interesting that we find a rather simple expression by using integral representation of Barnes' type. 
We expect that we can also obtain a similar expression for $X_8[1^2,2^3]$.\footnote{
$X_{12}[1^2,2^2,6]$ and $X_8[1^2,2^3]$ are both $2$-moduli models and have K3-fibration structure \cite{KV,KLM}. 
} 
Although we have correct coefficients for these examples, it is not clear whether we have the correct normalization for generic cases. 
However, these examples show that the direct integration via analytic continuation is a powerful method 
to get the correct form of the solutions of the inhomogeneous Picard-Fuchs equations.

Before closing this section we note that we have obtained the disk invariants for several other models by the same method. 
However it sometimes happens that the $Y$-coordinate dependence in the chain integration is not quadratic but cubic or a polynomial of higher degrees. 
It is likely that such complications are present in the situation where the weights of the ambient projective space are not all the same. 
Although we can obtain the integral disk invariants correctly, the computations and the interpretations seem to be a little bit ad-hoc.

\section{Application to the off-shell effective superpotential}
\subsection{Relative period integral}

Next we extend our analysis to the {\it off-shell formalism}, typically computations of superpotentials for toric branes 
\cite{JS1,JS2,AHMM,AHJMMS,GHKK,GHKK2,AB}. 
The research of the mirror symmetry of the toric brane in non-compact Calabi-Yau $3$-fold is initiated in \cite{AV,AKV}. 
The toric brane is specified by the extended toric charge vectors $\ell^{(a)}$.
One can construct the mirror pair of A- and B-brane from the extended
set of the toric vectors systematically. 

In \cite{JS1} it is proposed that the relative period of $H^3(X_3,S)$ 
for the holomorphic curve $S$ is assumed to be equivalent to that of 
$H^3(X_3,V)$ where $(S \subset ) \ V \subset X_3$ is a divisor, 
i.e. a complex codimension $1$ variety in the Calabi-Yau $3$-fold $X_3$ \footnote{
Mathematical justification is argued in \cite{LLY}. 
}. 
This aspect is studied further recently in relation with the non-compact Calabi-Yau $4$-fold \cite{AB}\footnote{ 
The relation of the relative period and the period for the non-compact $4$-fold is discussed in the appendix. } 
and F-theory on it \cite{AHJMMS,GHKK2}. 
The original open-closed duality of this kind is discussed in \cite{mayr}. 
The relation to the heterotic theory by the further duality chain is discussed in \cite{GHKK3,JMW}. 

The divisor $V$ is given by the single defining equation 
$Q(x_i,\phi)=0$ which is defined by one of the extended toric charge vector. 
For example in the case of the quintic $3$-fold, 
the defining equation for the divisor is given by 
\begin{equation}
Q(x_i,\phi)=x_{5}^5-\phi x_{1}x_{2}x_{3}x_{4}x_{5}\label{divisor_eq_quintic} 
\end{equation}
corresponding to the toric charge vector $\ell^{(2)}=(-1,0,0,0,0,1)$ \cite{AV}.\footnote{
We choose the toric vector for the quintic $3$-fold as $\ell^{(1)}=(-5,1,1,1,1,1)$. } 
Here we introduce the parameter $\phi$ and call it  open string moduli.

The chain integral $\int_{\Gamma}\Omega$ is regarded as the relative period integral. 
The holomorphic $3$-form $\Omega$ is extended to the relative $3$-form 
$\underline{\Xi}$ in the relative cohomology class $H^3(X_3,V)$. 
The relative $3$-form $\underline{\Xi}$ is a pair of the closed $3$-form 
$\Xi$ on $X_3$ and closed $2$-form $\xi$ on $V$, namely 
\begin{eqnarray}
\underline{\Xi}=(\Xi,\xi). 
\end{eqnarray}
Let $\iota$ be an embedding map 
\begin{eqnarray}
\iota :V \hookrightarrow X_3. 
\end{eqnarray}
The relative $3$-form $\underline{\Xi}\in H^3(X_3,V)$ satisfies the equivalence relation: 
\begin{eqnarray}
\underline{\Xi}\sim \underline{\Xi}+(d\alpha,\iota^*\alpha-d\beta), 
\end{eqnarray}
where $\alpha$ is a $2$-form on $X_3$ and $\beta$ is a $1$-form on $V$. 

The relative $3$-form $\underline{\Xi}$ is integrated over the relative $3$-cycle $\underline{\Gamma}\in H_3(X_3,V)$. 
The relative $3$-cycle is also decomposed as $\underline{\Gamma}=(\Gamma,\partial\Gamma)$. 
The relative $3$-forms and $3$-cycles are paired by the integration. 
\begin{eqnarray}
\int_{\underline{\Gamma}}\underline{\Xi}:=\int_{\Gamma}\Xi-\int_{\partial \Gamma}\xi. 
\end{eqnarray}
This definition is consistent with the above equivalence relations. 

The relative period is also evaluated by the Griffiths' residue integral formula \cite{Griffiths}. 
Now we assumed the original Calabi-Yau $3$-fold $X_3$ is a complete intersection in a weighted projective space $\mathbb{WP}^{n}$. 
$X_3$ is specified by $(n-3)$ defining equations $W_a \ (a=1,\cdots, n-3)$. 
For the embedding map $\iota:V\hookrightarrow X_3$, the pull back of a 
form $\alpha$ on $X_3$ is computed by inserting Poincar\'e residue map \cite{Voisin} 
\begin{eqnarray}
\iota^*\alpha=\frac{1}{(2\pi i)^2}\int_{T_{\epsilon}(V)}
\frac{dQ}{Q}\wedge \alpha. 
\end{eqnarray}
Then one finds that all the relative period integrals arise from 
a relative period $\Pi_r$ with a $\log Q$ factor.
\begin{eqnarray}
\Pi_r
=\int
\frac{\log Q(x,\phi)}{\prod_a W_a(x,\psi)}\Delta, 
\label{relative_int}
\end{eqnarray}
and $\Delta$ is a $n$-form on $\mathbb{WP}^n$ given by
\begin{eqnarray}
\Delta=\sum_{i=1}^{n+1}(-1)^{i+1}w_i dx^1\wedge \cdots\widehat{dx_i}\cdots \wedge dx^{n+1}, 
\end{eqnarray}
where $w_{i}$ are weights of $\mathbb{WP}^n$. 
This period satisfies the extended Picard-Fuchs equation. 
One of the solution which depends on the open string moduli $\phi$ is 
the effective superpotential $W_{\rm eff}(\psi,\phi)$. 
Extremizing this effective superpotential, one can fix the open string 
moduli as the function of the closed string moduli $\psi$. 
In fact, for the case of the quintic, the effective superpotential is extremized at $\phi=5\psi$. 
This logarithmic factor prescription is proposed first in \cite{JS1}. 
Next let us discuss the relation of the superpotentials in more detail. 

\subsection{Relative period and $4$-fold}
In the recent works \cite{AHJMMS,GHKK2,AB} it is discussed that 
the type IIB theory with D5-brane on Calabi-Yau $3$-fold $X_3$ is
related with 
F-theory on the non-compact Calabi-Yau $4$-fold $X_4$. 
The corresponding non-compact Calabi-Yau $4$-fold $X_4$ is given by the 
complete intersection of $W_a(x_i,\psi)=0$ and the defining equation 
\begin{eqnarray} 
Q_4(\phi)=x_{n+2} x_{n+3}+Q(x_i,\phi)=0, 
\end{eqnarray}
in the ambient space $\mathbb{WP}^{n+2}$ whose weights of coordinates $x_{n+2}$ and $x_{n+3}$ are determined 
by the defining equations.\footnote{
If the weight of $x_{n+2}$ or $x_{n+3}$ is zero, the resulting space becomes $\mathbb{WP}^{n+1}\times \mathbb{C}$.
} 
We also eliminate the points 
$(0:\cdots :0:x_{n+2}:x_{n+3})$ in $\mathbb{WP}^{n+2}$. 
The period of the holomorphic $4$-form on this geometry is 
\begin{eqnarray}
\Pi_4=\int_{T_\epsilon (\Gamma_4)}\frac{\Delta_{n+2}}{\prod_aW_a(x_i,\psi)Q_4(\phi)},
\end{eqnarray}
where $T_\epsilon (\Gamma_4)$ is the tubular neighborhood of the
$4$-cycle $\Gamma_4$ \cite{AB} and 
$\Delta_{n+2}$ is appropriate $(n+2)$-form on ambient space
$\mathbb{WP}^{n+2}$.

Therefore we have three different-looking formulas for $D5$-brane superpotentials: 
\begin{enumerate}
\item the chain integral of $3$-fold, 
\item the relative period of $3$-fold with the logarithmic factor, 
\item the period of $4$-fold. 
\end{enumerate}
Various analyses and discussions show that these three formulas are essentially the same things. 
Now we try to confirm this equivalence formally. 

We first discuss the connection between $4$-fold periods and logarithmic periods. 
In order to evaluate the $4$-fold period integral without suffering from the divergence, 
we consider its derivative with respect to the open string moduli $\phi$.
Here we change the variables $(x_{n+2},x_{n+3})$ to the polar coordinates such that 
\begin{eqnarray}
x_{n+2}=r e^{i\theta},\quad x_{n+3}=r e^{-i\theta}. 
\end{eqnarray}
The derivative of the period $\Pi_{4}$ is 
\begin{eqnarray}
\partial_{\phi} \Pi_{4}
&=&\int_{T_\epsilon (\Gamma_4)} \frac{\partial_{\phi}Q(x_i,\phi)\Delta_n\wedge d(r^2)\wedge d\theta}{\prod_a W_a(x_i,\psi)(r^2+Q)^2}. 
\end{eqnarray}
We define $n$-form $\Delta_n$ by the relation $\Delta_{n+2}=\Delta_n\wedge d(r^2)\wedge d\theta$. 
The integral over $r$ and $\theta$ can be performed by taking them as
the coordinates of the whole two dimensional plane. 
As a result, we find 
\begin{eqnarray}
\partial_{\phi} \Pi_{4}&=&
-\frac{1}{\pi}
\int_{T_\epsilon(\Gamma_3)} \frac{\partial_{\phi}Q(x_i,\phi)\Delta_n}{\prod_aW_a(x_i,\psi)Q(x_i,\phi)}
\nonumber \\
&=& -\frac{1}{\pi}\partial_{\phi}
\int_{T_\epsilon(\Gamma_3)} \frac{\log
Q(x_i,\phi)\Delta_n}{\prod_aW_a(x_i,\psi)}
\simeq \partial_{\phi}\Pi_r. \label{period_4-fold}
\end{eqnarray} 
Although this derivation is formal, we find that 
the $4$-fold period integral $\Pi_4$ is equivalent to
the relative period $\Pi_r$ up to the terms which do not 
depend on the open string moduli.

Now we discuss the relationship between the  
$4$-fold period integrals and the $3$-chain integrals on compact $3$-fold. 
We start with the formula in the first line of \eqref{period_4-fold}. 
Let $s$ be a local coordinate normal to the locus $Q(x_i,\phi)=0$ and
$\Delta_n = ds \wedge \Delta'$. 
One finds
\begin{eqnarray}
\partial_{\phi} \Pi_{4}&=&
-\frac{1}{\pi}
\int_{T_\epsilon(\Gamma_3)} \frac{\partial_{\phi}Q(x_i,\phi)ds\wedge\Delta'}{\prod_aW_a(x_i,\psi)Q(x_i,\phi)}
\nonumber \\
&=& -2i\int_{T_\epsilon(C)} \frac{\partial_{\phi}Q(x_i,\phi)\Delta'}{\prod_aW_a(x_i,\psi)\partial_{x_j}Q(x_i,\phi)\frac{dx_j(s)}{ds}}. 
\end{eqnarray}
The second equality follows by performing the residue integral with respect to $s$. 
Here we define the $2$-cycle $C=\Gamma_3\cap \{Q(x_i,\phi)=0\}$ and we
assume that this can be identified with the boundary of $3$-chain in $3$-fold. 

For the complete intersection Calabi-Yau $3$-fold, 
the three chain integral $\Pi_3$ takes the form
\begin{eqnarray}
\Pi_3=\int_{T_\epsilon (\Gamma_{\phi})}\frac{\Delta_n}{\prod_aW_a(x_i,\psi)}
\end{eqnarray}
where $T_\epsilon(\Gamma_{\phi})$ is the tubular domain around the $3$-chain $\Gamma_\phi$ 
whose boundary $\partial\Gamma_\phi=C_\phi$ is specified by
$Q(x_i,\phi)=0$.
Since the boundary deforms as 
\begin{eqnarray}
\partial_{x_j}Q\frac{dx_j(s)}{ds}ds+\partial_{\phi}Qd\phi=0, 
\end{eqnarray}
we have 
\begin{eqnarray}
\partial_\phi\Pi_3&=&\partial_\phi \int_{T_\epsilon(\Gamma_\phi)} \frac{ds\wedge\Delta'}{\prod_aW_a(x_i,\psi)}
=\int_{T_\epsilon(C_\phi)} \frac{\Delta'}{\prod_aW_a(x_i,\psi)}\frac{ds}{d\phi}
\nonumber\\
&\sim&\partial_\phi \Pi_4 
\end{eqnarray}
with identification $C=C_\phi$. 
This tells us that the chain integrals are directly related to the $4$-fold period integrals. 

Thus we formally realized the equivalence of the relative period with logarithmic differential for compact $3$-fold, 
the non-compact $4$-fold's period and the chain integral.\footnote{
The direct residue integral of $x_{n+2}$ and $x_{n+3}$ are easily 
performed, because the points $(0:\cdots :0:x_{n+2}:x_{n+3})$ are removed. 
The residue integral of $\oint dx_{n+3}/Q_4$ gives rise to a simple integral for $x_{n+2}$ coordinate: 
$\oint dx_{n+2}/x_{n+2}$. This residue integral picks up a point $x_{n+2}=0$, 
then we obtain the restricted $3$-fold period integral 
\begin{eqnarray}
\Pi_4\sim \int_{T_{\epsilon} (X_3)}\frac{\Delta}{\prod_aW_a(x_i,\psi)}\Biggl|_{Q(x_i,\phi)=0}. 
\end{eqnarray}
Such restriction can be rewritten as the insertion of the logarithmic 
form by the Poincar\'e residue map. 
Therefore we can check the validity of the above discussion. 
} 
In the following we will evaluate (\ref{relative_int}) 
directly via analytic continuation.

\subsection{Direct computation of relative period for mirror quintic}
Now we compute the relative period integral for the mirror 
quintic $3$-fold $Y_5$ via analytic continuation. 
The relative $3$-form is integrated over the tubular domain of the complete intersection of 
\begin{equation}
W(x_i,\psi)=\sum_{i=1}^5 x_{i}^5-5\psi x_{1}x_{2}x_{3}x_{4}x_{5}=0 
\end{equation}
and the divisor\footnote{
Of course other choices of divisor are possible. 
We choose one of the defining equations of curve and change one of coefficients into the open moduli. 
In \cite{AB}, the authors show that in their method all of results coincide each other 
at the critical locus of open moduli (on-shell), for the quintic case. 
} 
\begin{equation}
Q(x_i,\phi)=x_{5}^5-\phi x_{1}x_{2}x_{3}x_{4}x_{5}=0. 
\end{equation}
The complete intersection is also covered completely by two patches $x_1=1$ and $x_3=1$. 
Since the contribution to the residue integrals from these two patches
are the same, 
we multiply the factor two for the computation in $x_1=1$ patch.

In the $x_1=1$ patch, we can use the parametrization 
$(X,Y,Z,T)$ in the previous section \eqref{local coordinate quintic}. 
In terms of this parametrization, the relative period $\Pi_{r}$ is given by
\begin{eqnarray}
\Pi_{r}=2\cdot\int\frac{\omega_0\log [Y^{4/5}(1-\phi TXZ)]}{W}.
\end{eqnarray}

Here we are only interested in $\phi$-dependent term, so we neglect 
the term $\log Y^{4/5}$ in the numerator. 
Changing the parameters to polar ones, $\zeta$ and $w$, we can rewrite this integral 
\begin{eqnarray}
\Pi_{r}&=&2\cdot\frac{10}{(2\pi i)^4}\int\frac{d\zeta}{\zeta}\frac{dT}{T}wdwdY \nonumber\\
&&
\times \frac{\log (1-\frac{\phi}{5\psi}w^2)}{1-T^5+(5\psi)^{-5/2}(\zeta-\zeta^{-1})w^5T^{-5/2}Y^2+Y(1-w^2)}.
\end{eqnarray}
After the residue integral of $Y$ and the analytic continuation to the Barnes form, 
we obtain
\begin{eqnarray}
\Pi_{r}&=&2\cdot40\int\frac{ds}{2\pi i}\frac{\pi\cos(\pi s)}{\sin(\pi
 s)}\frac{\Gamma(2s+1)}{\Gamma(s+1)^2}z^{s/2}
\int dw\frac{w^{5s+1}\log(1-\frac{\phi}{5\psi}w^2)}{(1-w^2)^{2s+1}}
\nonumber \\
&&\times\int\frac{d\zeta}{2\pi i\zeta}(\zeta-\zeta^{-1})^s
\int\frac{dT}{(2\pi i)T}(T^{5/2}-T^{-5/2})^s.
\end{eqnarray}
The integrals of $\zeta$ and $T$ can be performed as before, so 
let's concentrate on the integration of $w$.

\begin{figure}[htbp]
\begin{center}
\includegraphics[width=10cm,angle=0,clip]{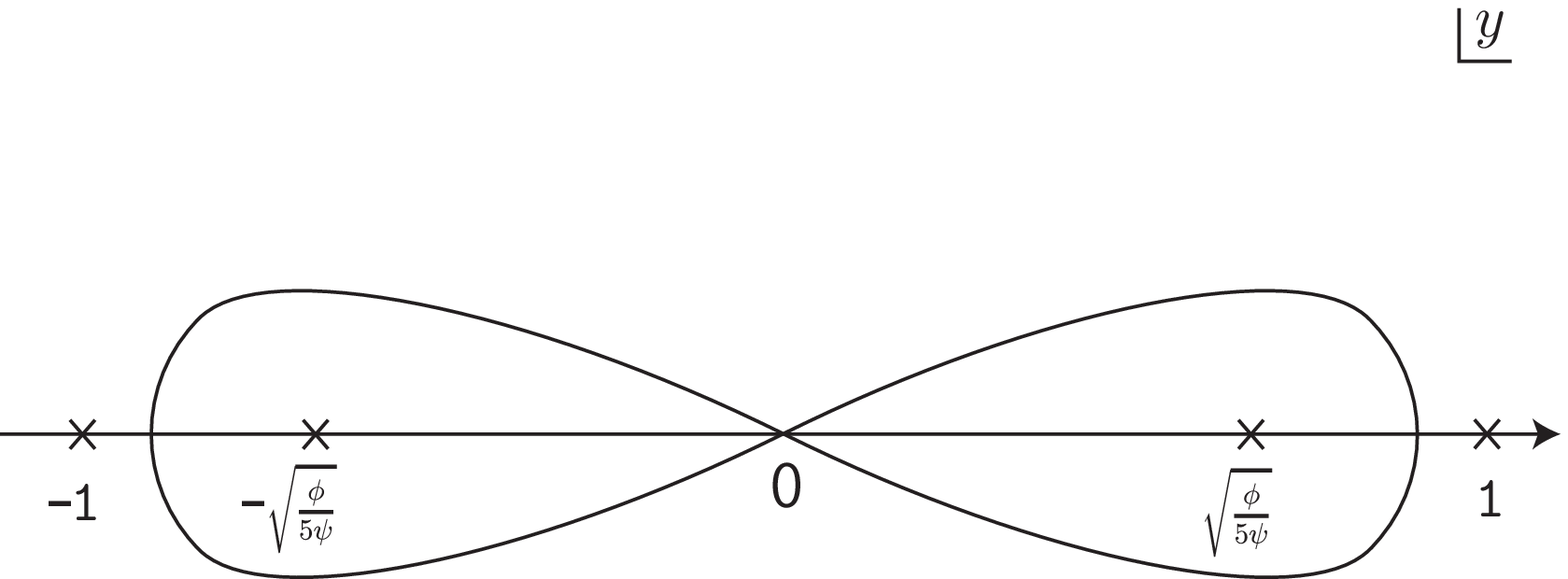}
\caption{Contour $C^{\prime}$}
\label{log_contour}
\end{center}
\end{figure}

To discuss the case of $\left|\frac{\phi}{5\psi}\right|<1$,
we change the integration variable to $y=1/w$.
\begin{eqnarray}
\frac{1}{2\pi i}\int_{C_w} dw 
\frac{w^{5s+1}\log (1-\frac{\phi}{5\psi}w^2)}{(1-w^2)^{2s+1}}
=\frac{1}{2\pi i}\int_{C_y} dy
\frac{\log\left(1-\frac{5\psi}{\phi}y^2\right)}{(y^2-1)^{2s+1}y^{s+1}},
\label{log_branch}
\end{eqnarray}
In the integrand, there exist logarithmic branch cuts which 
arises from the points $y=\pm\sqrt{\frac{\phi}{5\psi}}$.
We choose the contour $C_y$ surrounding $y=\pm
\sqrt{\frac{\phi}{5\psi}}$ as described in Fig.\ref{log_contour}.
Taking into account for the logarithmic branch, 
we can rewrite the integral $y$ as a integral over $[-\sqrt{\frac{\phi}{5\psi}},\sqrt{\frac{\phi}{5\psi}}]$.
\begin{eqnarray}
&&\frac{1}{2\pi i}\int_{C^{\prime}_y} dy
\frac{\log\left(1-\frac{5\psi}{\phi}y^2\right)}{(y^2-1)^{2s+1}y^{s+1}}
\nonumber \\
&&=\frac{1}{2\pi i}\int_{-\sqrt{\frac{\phi}{5\psi}}}^{\sqrt{\frac{\phi}{5\psi}}}
dy\frac{ \log\left(1-\frac{5\psi}{\phi}y^2\right)}{(y^2-1)^{2s+1}y^{s+1}}
+\frac{1}{2\pi i}
\int_{\sqrt{\frac{\phi}{5\psi}}}^{-\sqrt{\frac{\phi}{5\psi}}}dy
\frac{\log\left(1-\frac{5\psi}{\phi}y^2\right)-2\pi i}{(y^2-1)^{2s+1}y^{s+1}}
\nonumber \\
&&=\int_{-\sqrt{\frac{\phi}{5\psi}}}^{\sqrt{\frac{\phi}{5\psi}}}
dy \frac{1}{(y^2-1)^{2s+1}y^{s+1}}.
\end{eqnarray} 
Changing the integration variable to $x=y^2$, we find a simple expression.
\begin{eqnarray}
&&\int_{-\sqrt{\frac{\phi}{5\psi}}}^{\sqrt{\frac{\phi}{5\psi}}}
dy \frac{1}{(y^2-1)^{2s+1}y^{s+1}}
=\int_0^{\frac{\phi}{5\psi}}\frac{dx}{(x-1)^{2s+1}x^{\frac{s}{2}+1}}
\nonumber \\
&&=(-1)^{2s+1}\int_0^{\frac{\phi}{5\psi}}
\frac{dx}{(1-x)^{2s+1}x^{\frac{s}{2}+1}}.
\end{eqnarray}
Here we adopt a formula for the incomplete beta function
to evaluate the above integral. 
\begin{eqnarray}
B_z(p,q):&=&\int_0^zdt\;t^{p-1}(1-t)^{q-1}
=\frac{z^p}{p}F(p,1-q,p+1;z)\nonumber\\
&=&\frac{1}{\Gamma(1-q)}\sum_{\ell=0}^{\infty}
\frac{\Gamma(1-q+\ell)}{(p+\ell)\ell!}z^{\ell+p},
\end{eqnarray}
where ${\rm Re}\;z<1$ and the hypergeometric function is expanded as
\begin{eqnarray}
F(\alpha,\beta,\gamma;z)=\frac{\Gamma(\gamma)}{\Gamma(\alpha)\Gamma(\beta)}
\sum_{\ell=0}^{\infty}\frac{\Gamma(\alpha+\ell)\Gamma(\beta+\ell)}{\Gamma(\gamma+\ell)}
\frac{z^{\ell}}{\ell !}.
\end{eqnarray}
From this expression of the incomplete beta function, 
the $w$-integration can be expressed as $B_{\phi/5\psi}(p,q)$ with $p=-s/2$ and $q=-2s$. 
Finally we obtain 
\begin{eqnarray}
&&\frac{1}{2\pi i}\int_{C_w} dw \frac{w^{5s+1}\log
 (1-\frac{\phi}{5\psi}w^2)}{(1-w^2)^{2s+1}}
\nonumber \\
&&=(-1)^{2s+1}\frac{1}{\Gamma(2s+1)}\sum_{\ell=0}^{\infty}\frac{\Gamma(2s+1+\ell)}{\left(\ell-\frac{s}{2}\right)\ell
 !}\left(\frac{\phi}{5\psi}\right)^{-\frac{s}{2}+\ell}.
\end{eqnarray}
The integrals with respect to $\zeta$ and $T$ are given by
\eqref{zeta_int} and \eqref{T_int} as was the case of the on-shell chain
integral. 
As a result, the relative period becomes 
\begin{eqnarray}
\Pi_r&=&2\cdot \frac{1}{2\pi i}\int \frac{ds}{2\pi i}(5\psi)^{-5s/2}
\frac{\pi\cos(\pi s)}{\sin(\pi
s)}\frac{1}{\Gamma\left(\frac{s}{2}+1\right)^4}e^{\pi is/2}
2\pi \cos\left(\frac{\pi s}{2}\right)
\nonumber \\
&&\hspace*{0cm}
\times(-1)^{2s+1}\sum_{\ell=0}^{\infty}\frac{\Gamma(2s+1+\ell)}
{\left(\ell-\frac{s}{2}\right)\ell
!}\left(\frac{\phi}{5\psi}\right)^{-\frac{s}{2}+\ell}
\nonumber \\
&=&
\frac{2}{2\pi i}\int\frac{ds}{2\pi i}
\sum_{\ell=0}^{\infty}
\frac{2\pi^2\cos(\pi s)\cos(\frac{\pi s}{2})}{\sin(\pi s)}
\frac{\sin^4(-\frac{\pi s}{2})}{\pi^3\sin[\pi(-2s-\ell)]}
e^{\pi i s/2}(-1)^{2s+1}
\nonumber \\
&&\times
\frac{\Gamma(-\frac{s}{2})^4}{(\ell-\frac{s}{2})
\Gamma(-2s-\ell)\ell !}\phi^{-\frac{s}{2}+\ell}(5\psi)^{-\ell-2s},
\end{eqnarray}
where we used the reflection formula of the gamma function
\begin{eqnarray}
\Gamma(z)\Gamma(1-z)=\frac{\pi}{\sin (\pi z)}.
\end{eqnarray}
In the residue integral for $s$, there are simple poles at $s=-2k-1$ with positive integer $k$. 
Here we introduce a parameter $n:=\ell+k$. 
to rewrite the summation with respect to $\ell$, 
For $k\leq n\leq 5k+1$, the gamma function is finite. 
Picking up such poles, we find the relative period in the vicinity of the orbifold point as follows: 
\begin{eqnarray}
\Pi_{r}
&=&\frac{2}{\pi^2}\sum_{k=0}^{\infty}\sum_{n=k}^{5k+1}
\frac{(-1)^{n-k}\Gamma\left(k+\frac{1}{2}\right)^4}{(2n+1)(n-k)!(5k-n+1)!}\phi^{n+1/2}(5\psi)^{5k-n+2}.
\label{relative_quint}
\end{eqnarray}
This result coincides with the relative period which is obtained via extended Picard-Fuchs equation \cite{JS1}.\footnote{
Our normalization of the holomorphic $3$-form $\Omega$ differs from that
of \cite{JS1} by a factor $5\psi$. 
} 
Moreover, at the critical locus of the open moduli, $\phi\rightarrow 5\psi$, we recover on-shell situation: 
\begin{eqnarray}
\Pi_{r}\Big|_{(\phi)^{\frac{1}{2}}=\pm\sqrt{5\psi}}
&=&\frac{2}{\pi^2}\sum_{k=0}^{\infty}\sum_{n=k}^{5k+1}
\frac{(-1)^{n-k}\Gamma\left(k+\frac{1}{2}\right)^4}{(2n+1)(n-k)!(5k-n+1)!}(5\psi)^{5k+5/2}\nonumber\\ 
&=&\frac{1}{\pi^2}\sum_{k=0}^{\infty}\frac{\Gamma(k+\frac{1}{2})^5}{\Gamma(5k+\frac{5}{2})}(5\psi)^{5k+5/2}. \label{relative period quintic}
\end{eqnarray}
The second equality follows by the relation 
\begin{eqnarray}
\sum_{n=k}^{5k+1}\frac{(-1)^{n-k}}{(2n+1)(n-k)!(5k-n+1)!}
&=&\frac{1}{2}\sum_{l=0}^{4k+1}\frac{(-1)^l}{(k+l+\frac{1}{2})l!(4k-l+1)!}\nonum\\
&=&\frac{1}{2}\frac{\Gamma(k+\frac{1}{2})}{\Gamma(5k+\frac{5}{2})} 
\end{eqnarray}
and this can be obtained by the following formula: 
\begin{eqnarray}
\frac{\Gamma(x)}{\Gamma(x+a)}=\sum_{l=0}^\infty\frac{(-1)^l}{(x+l)\Gamma(l+1)\Gamma(a-l)}\label{gamma_formula}
\end{eqnarray}
with $x=k+1/2$ and $a=4k+2$.

Here in \eqref{relative period quintic}, the expression $(\phi)^{1/2}=\pm\sqrt{5\psi}$ means that $(\phi)^{1/2}=+\sqrt{5\psi}$ can be interpreted as $\Pi_+$ 
(the superpotential of $C_+$) 
and $(\phi)^{1/2}=-\sqrt{5\psi}$ as $\Pi_-$ (that of $C_-$) \cite{JS1}. 
Recall that the divisor equation $Q$ is given in \eqref{divisor_eq_quintic} and with $x_1+x_2=x_3+x_4=0$, 
\begin{align}
Q=x_{5}^5-\phi x_{1}x_{2}x_{3}x_{4}x_{5}\big|_{x_{1}+x_{2}=x_{3}+x_{4}=0}=x_5(x_5^2+(\phi)^{1/2}x_1x_3)(x_5^2-(\phi)^{1/2}x_1x_3). 
\end{align}
Comparing this to the defining equation of the curve in $C_\pm$ \eqref{curve_quintic} leads to the above statement. 
Here we consider the locus $\{x_5=0\}$ is irrelevant. 
The domainwall tension can be expressed by $\Pi_{r}\big|_{(\phi)^{1/2}=+\sqrt{5\psi}}-\Pi_{r}\big|_{(\phi)^{1/2}=-\sqrt{5\psi}}$ 
and it is nothing but \eqref{relative period quintic}.

\subsection{Mirror of double cubic $X_{3,3}[1^6]$}
We further apply our analytic continuation method for the mirror of the double cubic 
Calabi-Yau complete intersection $X_{3,3}[1^6]$. 
The defining equations for the mirror Calabi-Yau $3$-fold $Y_{3,3}$ are given by $W_1$ and $W_2$ in \eqref{double cubic def}. 
Now we consider the B-brane which is again given by the intersection with the hyperplanes $x_1+x_2=0$, $x_4+x_5=0$. 
Then, to introduce the open string moduli $\phi$, we will consider a divisor 
\begin{eqnarray}
Q(\phi):=x_{3}^3-\phi x_{4}x_{5}x_{6}=0. \label{cubic_divisor}
\end{eqnarray}

Now we compute the relative period $\Pi_{r}(\psi,\phi)$, 
\begin{eqnarray}
\Pi_{r}(\psi,\phi)=\frac{3^4\psi^2}{(2\pi i)^6}
\int\frac{\omega_0 \log Q}{W_1W_2}, 
\label{relative_cubic}
\end{eqnarray}
where $Q(x_i,\phi)$ is defined in (\ref{cubic_divisor}) and 
$\omega_0$ is a $5$-form on ${\mathbb{CP}}^5$ defined similarly as the quintic case. 

To evaluate this integral, we again consider in the $x_5=1$ patch and introduce the set of coordinates \eqref{double_cubic_1} 
and the polar ones \eqref{double_cubic_2}. 
Recall the coordinates for $x_2=1$ patch can be also found just by exchanging 
$(x_1,x_2,x_3)$ and $(x_4,x_5,x_6)$ and it is enough to parametrize the boundaries $C_\pm$ by these two partches. 
In terms of these coordinates, in the $x_5=1$ pacth, we can rewrite relative period integral 
(\ref{relative_cubic}) up to $\phi$ independent term in the following form:
\begin{eqnarray}
\Pi_{r}&=&\frac{2\cdot 6}{(2\pi i)^4}
\int\frac{w dw  dv dt d\zeta}{t\zeta}
\frac{\log(1-\frac{\phi}{3\psi}w^2)}{(1-w^2)(v-1)
-w^3v^{3/2}(\zeta-\zeta^{-1})(t^{3/2}-t^{-3/2})(3\psi)^{-3}}.
\nonumber \\
&=&\frac{2\cdot 6}{(2\pi i)^4}\sum_{n=0}^{\infty}z^{n/2}
\int\frac{w^{3n+1}\log(1-\frac{\phi}{3\psi}w^2)}{(1-w^2)^{2n+1}}
\int\frac{dv}{2\pi i}\frac{v^{3n/2}}{(v-1)^{n+1}}
\nonumber \\
&&\times \int\frac{d\zeta}{(2\pi i)\zeta}(\zeta-\zeta^{-1})^n
\int\frac{dt}{(2\pi i)t}(t^{3/2}-t^{-3/2})^n,
\end{eqnarray}
where we denote $z:=(3\psi)^{-6}$.

Here we consider the analytic continuation of the above period integral.
We replace the discrete sum with respect to $n$ by the contour integral
with respect to $s$.
The Barnes integral form of the relative period is 
\begin{eqnarray}
\Pi_{r}
&=&
\frac{2\cdot 6}{2\pi i}\int\frac{ds}{2\pi i}\frac{\pi \cos(\pi s)}{\sin(\pi s)}z^{s/2}
\int \frac{dw}{2\pi i}
\frac{w^{3s+1}\log(1-\frac{\phi}{3\psi}w^2)}{(1-w^2)^{s+1}}
\int \frac{dv}{2\pi i}
\frac{v^{3s/2}}{(v-1)^{s+1}}
\nonumber \\
&&\times \int\frac{d\zeta}{(2\pi i)\zeta}(\zeta-\zeta^{-1})^s
\int\frac{dt}{(2\pi i)t}(t^{3/2}-t^{-3/2})^s.
\end{eqnarray}

The integrals except $w$ can be performed in the same way as the on-shell formalism \eqref{double_cubic_integrals},
and the integral for $w$ is performed as in the above quintic case. 
We choose the same contour as Fig.\ref{log_contour}, 
and the result is 
\begin{eqnarray}
\int \frac{dw}{2\pi i}
\frac{w^{3s+1}\log(1-\frac{\phi}{3\psi}w^2)}{(1-w^2)^{s+1}}
=(-1)^{s+1}\frac{1}{\Gamma(s+1)}\sum_{\ell=0}^{\infty}\frac{\Gamma(s+1+\ell)}{(\ell-\frac{s}{2})\ell!}\left(\frac{\phi}{3\psi}\right)^{-\frac{s}{2}+\ell}.
\end{eqnarray}
We collect all results and perform $s$-integration. 
As a result of computations, the relative period in the vicinity of the orbifold point is found by picking up the poles at $s=-2k+1$ and becomes 
\begin{eqnarray}
\Pi_{r}=\frac{2}{\pi^2}\sum_{k=0}^{\infty}\sum_{n=k}^{3k}
\frac{(-1)^{n-k}\Gamma(k+\frac{1}{2})^5}{(2n+1)(n-k)!(3k-n)!\Gamma(3k+\frac{3}{2})}
\phi^{n+\frac{1}{2}}(3\psi)^{6k-n+\frac{5}{2}},
\label{relative_double_cubic}
\end{eqnarray}
where $\ell=n-k$. 

We also check our result in the limit $\phi\to 3\psi$ (the value at the critical locus of the open moduli $\phi$). 
In this limit, we find 
\begin{eqnarray}
&& \Pi_{r}\Big|_{(\phi)^{1/2}=\pm \sqrt{3\psi}}
=\frac{1}{\pi^2}\sum_{k=0}^{\infty}\frac{\Gamma(k+\frac{1}{2})^6}{\Gamma(3k+\frac{3}{2})^2}(3\psi)^{6k+3}.\label{relative period double}
\end{eqnarray}
In the derivation, as in the case of the quintic, we used the following identity: 
\begin{eqnarray}
\sum_{n=k}^{3k}\frac{(-1)^{n-k}}{(2n+1)(n-k)!(3k-n)!}=\frac{1}{2}\frac{\Gamma(k+\frac{1}{2})}{\Gamma(3k+\frac{3}{2})}.
\end{eqnarray}
This can be obtained from the formula \eqref{gamma_formula} with $x=k+1/2$ and $a=2k+1$. 

As discussed in the quintic case, 
the choice $(\phi)^{1/2}=\pm\sqrt{3\psi}$ can be interpreted as $\Pi_{\pm}$.
The domainwall tension can be obtained by $\Pi_{r}\big|_{(\phi)^{1/2}=+\sqrt{3\psi}}-\Pi_{r}\big|_{(\phi)^{1/2}=-\sqrt{3\psi}}$.

\subsection{Extended Picard-Fuchs equation: Consistency check}
As a consistency check of the above results, we discuss the extended Picard-Fuchs equation.
It is a homogeneous differential equation and 
its solutions are relative periods, which depends on both open and closed moduli. 
For toric Calabi-Yau $3$-fold, one can find the Picard-Fuchs equation from the toric data directly. 
Adding D-brane, we should consider the extended Picard-Fuchs system with the additional toric charge. 
The application of this method to the compact Calabi-Yau is discussed in \cite{AHMM}. 
See also \cite{AHJMMS,GHKK2} for further discussions. 

For a set of toric charge vectors $\ell^{(a)}=(\ell^{a}_i)$, we find the 
Picard-Fuchs operators ${\cal L}_a$ as follows: 
\begin{equation}
{\cal L}_a=\prod_{k=1}^{\ell_0^a}(\theta_{a_0}-k)
\prod_{\ell_i^a>0}\prod_{k=0}^{\ell_i^a-1}(\theta_{a_i}-k)
-(-1)^{\ell_0^a}z_a\prod_{k=1}^{-\ell_0^a}(\theta_{a_0}-k)
\prod_{\ell_i^a<0}\prod_{k=0}^{-\ell_i^a-1}(\theta_{a_i}-k), \nonumber
\end{equation}
\begin{equation}
z_a:=(-1)^{\ell_0^a}\prod_{i}c_i^{\ell_i^a},\quad 
\theta_{a_i}:=\sum_{a}\ell_i^a\theta_{z_a}, 
\end{equation}
where $c_i$ are coefficients of the defining equation $W=\sum_i c_iy_i$ 
for the mirror Calabi-Yau $3$-fold, and 
$\theta_{z_a}=z_a\partial/\partial z_a$. 

For the quintic Calabi-Yau $3$-fold case, as previously mentioned we choose the toric vectors as $\ell^{(1)}=(-5,1,1,1,1,1)$ 
and for divisor $Q(x_i,\phi)=0$ expressing B-brane as $\ell^{(2)}=(-1,0,0,0,0,1)$ \cite{AV}. 
One of the linear combinations of the toric charge vectors leads to a Picard-Fuchs operator that 
gives a nontrivial constraint on the relative period \cite{AHMM}
\begin{equation}
{\cal L}_1^{\prime}=\theta_2\theta_1^4+z_1z_2\prod_{i=1}^5(4\theta_1+\theta_2+i),
\end{equation}
where 
\begin{equation}
z_1=(5\psi)^{-4}\phi^{-1},\quad z_2=-(5\psi)^{-1}\phi.
\end{equation}

The toric charge vector for the double cubic complete intersection Calabi-Yau $3$-fold is
\begin{eqnarray}
\ell_1=(-3,-3,1,1,1,1,1,1). \label{toric_dc_1}
\end{eqnarray}
From this toric charge, we find the defining equations 
$W_1$ and $W_2$ in \eqref{double cubic def} for the mirror Calabi-Yau $3$-fold $Y_{3,3}$. 
There are four toric charge vectors to determine the B-brane locus, which is expressed as a curve. 
Now we consider the B-brane which is again given by the intersection with the hyperplanes $\{x_1+x_2=0\}$ and $\{x_4+x_5=0\}$. 
Then, to introduce the open string moduli $\phi$, we will consider a divisor 
\begin{eqnarray}
Q(\phi):=x_{3}^3-\phi x_{4}x_{5}x_{6}=0, 
\end{eqnarray}
defined by the extra toric charge vector 
\begin{eqnarray}
\ell_2=(-1,0,0,0,1,0,0,0). 
\label{toric_dc_2}
\end{eqnarray}
Toric charge vectors $\ell_1$ and $\ell_2$ give an extended Picard-Fuchs operator. 
From one of the linear combinations, 
we find an extended Picard-Fuchs operator 
\begin{eqnarray}
{\cal L}_1^{\prime}=\theta_2\theta_1^5-z_1z_2
\prod_{i=1}^3(2\theta_1+\theta_2+i)\prod_{i=1}^3(3\theta_1+i),
\label{PF_op_double}
\end{eqnarray}
where 
\begin{equation}
z_1=(3\psi)^5\phi^{-1},\quad z_2=(3\psi)^{-1}\phi.
\end{equation}

One can verify that in each case the relative periods \eqref{relative_quint} and \eqref{relative_double_cubic} 
satisfy the extended Picard-Fuchs equations 
\begin{equation}
{\cal L}^{\prime}_1\Pi_{r}(\psi,\phi)=0.
\end{equation} 
Thus we can confirm that the relative periods obtained via direct integration are surely solutions for the 
extended Picard-Fuchs equations.

\section{Normalization ambiguity}
In section $3$ and $4$, we have directly evaluated the D-brane superpotentials by use of analytic continuations. 
This method is a powerful approach for obtaining the analytical expressions of the superpotentials for both on-shell and off-shell. 
The disadvantage of this approach is that we cannot determine the normalization of the superpotentials 
since we may have ambiguities in the analytic continuations. 
Therefore, in this section we are going to try to fix the normalization by considering genus one amplitudes, focusing on the on-shell (i.e. involution brane) situations. 

The holomorphic anomaly equation connects the tree level amplitudes with the amplitudes of the higher worldsheet topologies \cite{BCOV1,BCOV2}. 
The extension of this to the open string sector, in particular for the compact Calabi-Yau, is proposed in \cite{Wa2,Wa3}. 
The related works are discussed in \cite{COY,BoTa}. 
We will use the formula for one loop amplitudes given in these works. 
Now we compute real BPS invariants at Euler characteristic $-2$, $0$ (genus $0$, $1$) 
for one parameter Calabi-Yau complete intersections in 
weighted projective spaces under the assumption 
that the formula for Euler characteristic $0$ obtained in \cite{Wa3} holds. 

In general we consider $X_{d_1,d_2,\cdots,d_k}(w_1,w_2,\cdots,w_l)$, 
a complete intersection of $k$ hypersurfaces of degrees $d_1$, $\cdots$, $d_k$ in the weighted projective space $\mathbb{WP}^{l-1}(w_1,\cdots,w_l)$.\footnote{
All of few moduli Calabi-Yau complete intersections in the weighted projective space are listed in \cite{table}. 
We here concern the following specific models: 
$X_5(1^5)$, $X_6(1^4,2)$, $X_8(1^4,4)$, $X_{10}(1^3,2,5)$, $X_{3,3}(1^6)$, $X_{4,4}(1^4,2^2)$, $X_{6,6}(1^2,2^2,3^2)$, $X_{4,3}(1^5,2)$. 
All these models are (closed) one-moduli models. 
See also \cite{Mo,KT1,KT2,LibTeit,KS,vEvS}. 
We can apply our direct integration method to these models and obtain the same enumerative numbers as \cite{KW1,KS1,Walcher} 
although there are subtleties we have mentioned already in the footnote of the section $3$. 
} 
Then the various ingredients are listed as follows. 
First, the condition of $3$-fold obtained by the adjunction formula is given by $l-k=4$. 
The classical Yukawa coupling (triple intersection number) is 
\bea
K_0=\frac{\prod_{i=1}^k d_i}{\prod_{i=1}^l w_i}. 
\ena
The first and second Chern class, $c_1$, $c_2$, and the Euler characteristic $\chi$ can be obtained as 
\begin{equation}
c_1=\sum_{i=1}^k d_i-\sum_{i=1}^l w_i=0, 
\end{equation}
\begin{equation}
c_2=\frac{1}{2}\frac{\prod_{i=1}^l d_i}{\prod_{i=1}^l w_i}\left(\sum_{i=1}^k d_i^2-\sum_{i=1}^l w_i^2\right), 
\end{equation}
\begin{equation}
\chi =\frac{1}{3}\frac{\prod_{i=1}^k d_i}{\prod_{i=1}^l w_i}\left(-\sum_{i=1}^k d_i^3+\sum_{i=1}^l w_i^3\right). 
\end{equation}
The discriminant of one-moduli models is
\bea
(diss)=\left(1-\frac{\prod_{i=1}^k {d_i}^{d_i}}{\prod_{i=1}^l {w_i}^{w_i}}z\right).
\ena
The fundamental period $\varpi_0$ and logarithmic period $\varpi_1$ are
\begin{equation}
\varpi_0=\sum_{n=0}^\infty \frac{\prod_{i=1}^k \Gamma(d_in+1)}{\prod_{i=1}^l \Gamma(w_in+1)}z^n,
\end{equation}
\begin{equation}
\varpi_1=\sum_{n=0}^\infty \frac{\prod_{i=1}^k \Gamma(d_in+1)}{\prod_{i=1}^l \Gamma(w_in+1)}\left[\log z +\left(\sum_{i=1}^k d_i \Psi(d_in+1)-\sum_{i=1}^l w_i \Psi(w_in+1)\right)\right]z^n 
\end{equation}
where $\Psi$ denotes the digamma function. 

It has turned out that the chain integral (i.e. the tension of the BPS
domainwall) is the following general form (for the $\mathbb{Z}_2$ vacua case\footnote{
The generalization to the $\mathbb{Z}_k$-sector $(k\neq 2)$ is easy and the formula in such case is in \cite{Walcher}. 
}) 
\bea
\Pi(z)(=\varpi_0(z)\mathcal{T}_A(z))=\frac{c}{2\pi^2}\sum_{n=0}^\infty \frac{\prod_{i=1}^l \Gamma(-w_in-\frac{w_i}{2})}{\prod_{i=1}^k \Gamma(-d_in-\frac{d_i}{2})}z^{n+\frac{1}{2}}, 
\ena
where $\mathcal{T}_A$ is the A-model domainwall tension (i.e. the disk generating function) under the transformation by the mirror map. 
The integer constant $c$ is the normalization factor and assumed to be given by $c=N/\vert \mathcal{S}\vert$. 
Here, $N$ is the number of branes which transforms under discrete group of the models and $\mathcal{S}$ is the stabilizer of the curves, 
namely, $\vert \mathcal{S} \vert$ counts the order of discrete subgroup which makes the brane invariant. 
This constant is in some way related to the constant factor of the inhomogeneous term of the Picard-Fuchs equation in \cite{Walcher}. 
Since we do not have any A-model calculus except for $X_5(1^5)$ and $X_{3,3}(1^6)$, we do not know the true normalization. 
In the following, we will find that the most of models have $c=1$. 
We will also note that the consistency and discrepancy with Walcher's results for some CICY models \cite{Walcher}.

There are other constraints for the weight since they must be one-moduli models. 
At the level of the closed string, the periods satisfy the fourth order homogeneous equation. 
Therefore, the fundamental period should have the form 
\bea
\varpi_0&=&\sum_{n=0}^\infty\frac{\prod_{i=1}^k\prod_{j=1}^{d_i}\frac{\Gamma(n+\frac{j}{d_i})}{\Gamma(\frac{j}{d_i})}}
{\prod_{i=1}^l\prod_{j=1}^{w_i}\frac{\Gamma(n+\frac{j}{w_i})}{\Gamma(\frac{j}{w_i})}}\left(\frac{\prod_{i=1}^k d_i^{d_i}}{\prod_{i=1}^l w_i^{w_i}}z\right)^n\nonumber\\
&=&\sum_{n=0}^\infty\frac{\Gamma(n+\lambda_1)\Gamma(n+1-\lambda_1)\Gamma(n+\lambda_2)\Gamma(n+1-\lambda_2)}{\Gamma(\lambda_1)\Gamma(1-\lambda_1)\Gamma(\lambda_2)\Gamma(1-\lambda_2)\Gamma(n+1)^4}
\left(\frac{\prod_{i=1}^k d_i^{d_i}}{\prod_{i=1}^l w_i^{w_i}}z\right)^n,
\ena
where $\lambda_1$, $\lambda_2$ are determined by 
\bea
\lbrace\lambda_1,1-\lambda_1,\lambda_2,1-\lambda_2\rbrace
&=&\left\{ \ \frac{j}{w_i} \ \Biggl|j=1,...,w_i-1, \ i=1,...,l\right\}\nonumber\\
&{}&-\left\{ \ \frac{j}{d_i} \ \Biggl|j=1,...,d_i-1, \ i=1,...,k\right\}.
\ena
We can rewrite the logarithmic period in terms of $\lambda_1$ and $\lambda_2$ as follows: 
\bea
\varpi_1&=&\varpi_0\log z+\tilde{\varpi}_1, 
\ena
where $\tilde{\varpi}_1$ is expressed as 
\bea
\tilde{\varpi}_1&=&\sum_{n=0}^\infty\frac{\Gamma(n+\lambda_1)\Gamma(n+1-\lambda_1)\Gamma(n+\lambda_2)\Gamma(n+1-\lambda_2)}{\Gamma(\lambda_1)\Gamma(1-\lambda_1)\Gamma(\lambda_2)\Gamma(1-\lambda_2)\Gamma(n+1)^4}\nonumber\\
&\times&\left[\sum_{i=1,2}\Psi(n+\lambda_i)-\Psi(\lambda_i)+\Psi(n+1-\lambda_i)-\Psi(1-\lambda_i)-4(\Psi(n+1 )-\Psi(1))\right]\nonumber\\
&{}&\times\left(\frac{\prod_{i=1}^k {d_{i}}^{d_i}}{\prod_{i=1}^l {w_{i}}^{w_i}}z\right)^n.
\ena
The mirror map is given by $t(z)=\frac{1}{2\pi i}\frac{\varpi_1(z)}{\varpi_0(z)}$. 
Under this map the genus zero (disk) amplitude in the A-model, $\mathcal{T}_A(z)$, which is given by 
\begin{equation}
\Pi(z)=\varpi_0(z)\mathcal{T}_A(z)=\frac{c}{2\pi^2}\sum_{n=0}^{\infty}
  \frac{\prod_{i=1}^l\Gamma(-nw_i-\frac{w_i}{2})}
 {\prod_{i=1}^k \Gamma(-nd_i-\frac{d_i}{2})}\, z^{n+\frac{1}{2}}, 
\end{equation}
has the enumerate structure in the A-model interpretation and 
we define {\it the genus $0$ real BPS invariants}, $n_d^{(0,\rm{real})}$ (for $d$ odd), by
\begin{equation}
\mathcal{T}_A(z)=\sum_{k,d:\textrm{ odd}}\frac{2n_d^{(0,\rm{real})}}{k^2}q^{kd/2},
\end{equation}
where $q=e^{2\pi it(z)}$ \cite{OV,Wa1,PaSoWa}. 
Here $d$ is the degree and $k$ is the integer for re-summation. 
Note that $\tau(z)$ is a solution to the inhomogeneous Picard-Fuchs equation 
$\mathcal{L}_{PF}\tau(z)=\frac{a}{2^4} z^{1/2}$ with 
\begin{equation}
a=2n_1^{(0,\rm{real})}
=c\frac{\Gamma(-\frac{w_1}{2})\cdots \Gamma(-\frac{w_l}{2})}
{2\pi^2\Gamma(-\frac{d_1}{2})\cdots \Gamma(-\frac{d_k}{2})}. 
\end{equation}

According to \cite{Wa3}, the holomorphic limits $\mathcal{A}^{\rm{hol}}$, 
$\mathcal{K}^{\rm{hol}}$ of 
the annulus amplitude $\mathcal{A}$ and the Klein bottle amplitude $\mathcal{K}$ are given by 
\cite[eq.(5.27)]{Wa3}\cite{KW} 
\begin{equation}
\frac{\partial}{\partial z}\mathcal{A}^{\rm{hol}}
=-\frac{1}{2}(\Delta_{zz}^{\rm{hol}})^2 C_{zzz}^{-1}, 
\qquad
\mathcal{K}^{\rm{hol}}=
\frac{1}{2}\log{\left[\frac{q}{z}\frac{dz}{dq}(diss)^{-1/4}\right]}. 
\end{equation}
Here $\Delta_{zz}^{\rm{hol}}$ is given by 
\begin{equation}
\Delta_{zz}^{\rm{hol}}=(\partial_z-\Gamma_{zz}^z+\partial_z K)
(\partial_z+\partial_z K)\tau(z), 
\end{equation}
where 
$\Gamma_{zz}^z=\partial_z\log\frac{dt(z)}{dz}$ 
and $\partial_z K=-\partial_z \log \varpi_0$, and 
the Yukawa coupling $C_{zzz}$ takes the form 
\begin{equation}
C_{zzz}=\frac{K_0}{z^3 \cdot (diss)}. 
\end{equation}
Another form of $\Delta_{zz}^{\rm{hol}}$ is given in terms of $q$ as 
\bea
\Delta_{zz}^{\rm{hol}} =\varpi_0\left(\frac{dz}{dq}\right)^{-2}\left(\partial_q+\frac{1}{q}\right)\partial_q \varpi_\Gamma. 
\ena
Then, 
{\it the genus $1$ real BPS numbers} $n_d^{(1,\rm{real})}$ are defined by 
the following expansion \cite[eq.(5.28)]{Wa3}: 
\begin{equation}
\mathcal{A}^{\rm{hol}}+\mathcal{K}^{\rm{hol}}
=\sum_{\begin{subarray}{c}
d \in 2\mathbb{Z}_{>0}, \\k\in2\mathbb{Z}_{\geq 0}+1
\end{subarray}
}\frac{2n_d^{(1,\rm{real})}}{k}q^{dk/2}. 
\label{genusoneformula}
\end{equation}

Because of the relation of the mirror map and 
the expression of the logarithmic periods in terms of $\lambda_1$, $\lambda_2$, we obtain the following formula: 
\bea
q&=&\exp{\left(\frac{\varpi_1}{\varpi_0}\right)}=z \exp({\tilde{\varpi}_1})\nonumber\\
&=&z+\lambda_1(1-\lambda_1)\lambda_2(1-\lambda_2)\left[\frac{1}{\lambda_1}+\frac{1}{1-\lambda_1}+\frac{1}{\lambda_2}+\frac{1}{1-\lambda_2}-4\right]\frac{\prod_{i=1}^k d_i^{d_i}}{\prod_{i=1}^l w_i^{w_i}}z^2+\cdots. \nonum\\
\ena
We can invert this as 
\bea
z=q-[\lambda_1(1-\lambda_1)+\lambda_2(1-\lambda_2)-4\lambda_1(1-\lambda_1)\lambda_2(1-\lambda_2)]\frac{\prod_{i=1}^k d_i^{d_i}}{\prod_{i=1}^l w_i^{w_i}}q^2+\cdots.
\ena
Substituting the expansion in (\ref{genusoneformula}), we have 
\bea
\mathcal{K}&=&\left[-\frac{1}{2}(\lambda_1(1-\lambda_1)+\lambda_2(1-\lambda_2)-4\lambda_1(1-\lambda_1)\lambda_2(1-\lambda_2))+\frac{1}{8}\right]\frac{\prod_{i=1}^k d_i^{d_i}}{\prod_{i=1}^lw_i^{w_i}}q+\cdots,\nonumber\\
&=&2\left(\lambda_1-\frac{1}{2}\right)^2\left(\lambda_2-\frac{1}{2}\right)^2\frac{\prod_{i=1}^{k} d_i^{d_i}}{\prod_{i=1}^lw_i^{w_i}}q+\cdots, 
\ena
and we can rewrite it as 
\bea
\mathcal{K}=2^{-7}\pi^{-4}\frac{\prod_{i=1}^l\Gamma(-\frac{w_i}{2})^2}{\prod_{i=1}^k\Gamma(-\frac{d_i}{2})^2}\frac{\prod_{w_i;\mathrm{odd}}w_i}{\prod_{d_i;\mathrm{odd}}d_i}q+\cdots.
\ena
On the other hand, we have
\bea
\mathcal{A}=-2^7\pi^{-4}\frac{\prod_{i=1}^l\Gamma(-\frac{w_i}{2})^2}{\prod_{i=1}^l\Gamma(-\frac{d_i}{2})^2}\frac{\prod_{i=1}^lw_i}{\prod_{i=1}^{k}{d_i}}c^2q+\cdots. 
\ena
Therefore, we finally get
\bea
n_2^{(1,\mathrm{real})}=2^{-8}\pi^{-4}\frac{\prod_{i=1}^l\Gamma(-\frac{w_i}{2})^2}{\prod_{i=1}^k\Gamma(-\frac{d_i}{2})^2}\left[\frac{\prod_{w_i;\mathrm{odd}}w_i}{\prod_{d_i;\mathrm{odd}}d_i}-\frac{\prod_{i=1}^lw_i}{\prod_{i=1}^{k}{d_i}}c^2\right].
\ena
In most cases, since we have $c=1$, we can find 
\begin{equation}
n_2^{(1,\mathrm{real})}=0, \ 24, \ 72, \ 2^{12}, \ 0, 
\end{equation}
for $X_5(1^5)$, $X_6(1^4,2)$, $X_8(1^4,4)$, $X_{10}(1^3,2,5)$, $X_{3,3}(1^6)$, respectively. 
For (the $\mathbb{Z}_2$-sector of) $X_{4,4}(1^4,2^2)$, we have $n_2^{(1,\mathrm{real})}=12$ if we assume $c=1$ and $n_2^{(1,\mathrm{real})}=0$ when $c=2$. 
The tree level result obtained by Walcher for $X_{4,4}(1^4,2^2)$ corresponds to $c=4$, from which we obtain negative integers for $n_2^{(1,\mathrm{real})}$. 
This result does not mean that $c=4$ for $X_{4,4}(1^4,2^2)$ is not a good choice because we have holomorphic ambiguity. 
Moreover, we have $n_2^{(1,\mathrm{real})}=3/2$ for $X_{4,3}(1^5,2)$ if we take $c=1$. 
Furthermore, we cannot find any good rational values for $c$ such that we have positive integers for $n_2^{(1,\mathrm{real})}$. 
This result may imply that we have to add some other terms for genus one amplitudes by considering holomorphic ambiguities.

\section{Conclusions}
In this paper we discuss new methods for open mirror symmetry of the compact Calabi-Yau $3$-fold. 
First, we have shown that the inhomogeneous Picard-Fuchs equation of the
chain integral can be obtained by a rather systematic algorithm. 
This algorithm is also effective for various CICYs. In such cases, the
Griffiths-Dwork method is not so easy task. 
Then, we have evaluated the superpotentials or domainwall tensions directly via
analytic continuations. 
We found that our method is a powerful approach for obtaining the analytical expressions of the superpotentials for both on-shell and off-shell formalism. 
We treat several models with a few moduli and reproduce the known results. 
The disadvantage of this approach is the problem of fixing the
normalization of the $3$-chain integral which may result 
from the ambiguity of analytic continuation.
So as to overcome this point, we have considered the genus one
amplitudes, and fixed the normalization appropriately. 
Our direct integration method might not be mathematically rigorous, 
but we can reproduce the known results rather easily 
and the computation is economical and intuitive. 
It is important that we can directly obtain the domainwall tension itself without treating any other relative periods. 
So far there is no result of compact Calabi-Yau manifolds except for a few (bulk) moduli hypersurface/CICY models\footnote{
While this paper was in preparation for submission, a related work appeared \cite{GKZ} and the analysis of open mirror symmetry on compact Calabi-Yau hypersurfaces with $2$- and $3$-moduli were carried out by very systematic toric and GKZ-system approach. 
}. 
We expect that our methods works for more general class of Calabi-Yau
$3$-fold as the Pffaffian Calabi-Yau varieties \cite{Kanazawa}, which will be reported elsewhere \cite{ShimizuSuzuki}. 
\\

\noindent{\bf Acknowledgements}
\\
We would like to thank Masao Jinzenji, Yukiko Konishi and Satoshi Minabe for discussions. 
One of the authors (H.F.) would like to thank Johannes Walcher for fruitful discussions, 
one of the authors (H.S.) would like to thank Daniel Krefl for stimulating discussions and 
one of the authors (M.S.) would like to thank Eric Gigu$\grave{\rm{e}}$re for helpful comments. 
H.F. is supported by the Grant-in-Aid
for Nagoya University Global COE Program, 
``Quest for Fundamental Principles in the Universe: from Particles 
to the Solar System and the Cosmos'', and the Grant-in-Aid for Young Scientists (B) [\# 21740179] from the Japan Ministry of Education, Culture, Sports, Science and Technology.
H.S. is supported by a Grant-in-Aid for Scientific Research on Priority Area 
(Progress in Elementary Particle Physics of the  21st Century through Discoveries of Higgs Boson and Super- symmetry, Grant No. 16081201) 
provided by the Ministry of Education, Science, Sports and Culture, Japan. 

\newpage

\appendix
\section{Rescaling algorithm and the inhomogeneous terms}\label{details}
In this appendix, we will mainly discuss the rescaling algorithm for
some examples.
\subsection{Cubic curve}
As the first example, we will consider the family of Calabi-Yau 1-fold (elliptic curve) defined as a hypersurface by the following homogeneous polynomial of degree three in $\mathbb{CP}^{2}$,
 \begin{eqnarray}
  W=\fr{1}{3}(x_{1}^{3}+x_{2}^{3}+x_{3}^{3})-\psi x_{1}x_{2}x_{3}=0. \label{torus}
 \end{eqnarray}
In this case we define the integral of two form over $2$-chain as 
 \begin{eqnarray}
  \int_{\Gamma}\Omega=\psi\int_{\Gamma}\fr{\omega_{0}}{W} 
 \end{eqnarray}
where
 \begin{eqnarray}
  \omega_{0}&=&\sum_{i=1}^{3}(-1)^{i}x_{i}dx_{1}\wedge\cdots\wedge\wh{dx_{i}}\wedge\cdots\wedge dx_{3}\nonum\\
            &=&-x_{3}dx_{1}\wedge dx_{2}+x_{2}dx_{1}\wedge dx_{3}-x_{1}dx_{2}\wedge dx_{3}
 \end{eqnarray}
and $\Gamma$ has $\psi$-dependence. 
For simplicity, we fix here one of the homogeneous coordinates of $\mathbb{CP}^{2}$, $x_1$, to $1$, 
so that $\omega_{0}=-dx_{2}\wedge dx_{3}$. 

We find from $\eqref{torus}$ (or precisely a redundant moduli version $W'$) the obvious differential relation  
 \begin{eqnarray}
  \prod_{i}^{3}\le(\fr{\p}{\p a_{i}}\ri)\fr{\omega_{0}}{W'}=\le(\fr{\p}{\p a_{0}}\ri)^{3}\fr{\omega_{0}}{W'},
 \end{eqnarray}
and reproduce this equation with $\psi$ derivatives in the similar way to the quintic case. 
To obtain the Picard-Fuchs equation which is second order, we start from the
following $\theta_{\psi}:=\psi\partial_\psi$ factorized form,
 \begin{eqnarray}
  \int\Omega&=&\psi\int\fr{-dx_{2}\wedge dx_{3}}{W} \nonum\\
  &=&\theta_{\psi}\int\log W(d\log x_{2}\wedge d\log x_{3}) \nonum\\
  &=&\theta_{\psi}\int\log W(d\log x_{2}\wedge d\log \t{x}_{3}) \nonum\\
  &&\le(x_{3}=\frac{\t{x}_{3}}{\psi},\quad W=\fr{1}{3}\left(x_{1}^{3}+x_{2}^{3}+\frac{\t{x}_{3}^{3}}{\psi^{3}}-3x_{1}x_{2}\t{x}_{3}\right)\ri) \nonum\\
  &=&\int\fr{x_{3}^{3}}{W}\le(-d\log x_{2}\wedge d\log x_{3}\ri)
   +\int\fr{\p (\log W)}{\p x_{3}}d\log x_{2}\wedge dx_{3}. \label{example}
 \end{eqnarray}
Here we used the following formula for the $\psi$ derivatives of the integration which has a boundary,
 \begin{eqnarray}
  \theta_{\psi}\int_{a}^{b}\fr{f(x)}{x}dx
  &=&\theta_{\psi}\int_{a/\psi^{m}}^{b/\psi^{m}}\fr{f(\psi^{m}\t{x})}{\t{x}}d\t{x} \nonum\\
  &=&\int_{a/\psi^{m}}^{b/\psi^{m}} \theta_{\psi}\fr{f(\psi^{m}\t{x})}{\t{x}}d\t{x}
   -m\int_{a}^{b}\fr{d}{d x}f(x)dx,
 \end{eqnarray}
where $x=\psi^{m}\t{x}$.

It is also easy to find the differential relation using this form of $\Omega$. 
Next, we consider reproducing the $\p/\p a_{1}$ derivative of \eqref{example}, using coordinate transformation,
 \begin{eqnarray}
  -\theta_{\psi}\int\fr{x_{3}^{3}}{W}d\log x_{2}\wedge d\log x_{3}
  &=&-\theta_{\psi}\int\fr{\t{x}_{3}^{3}}{\wt{W}}d\log x_{2}\wedge d\log x_{3} \nonum\\
  &&\le(x_{2}=\psi\t{x}_{2},\quad x_{3}=\psi\t{x}_{3},\quad \wt{W}=\fr{1}{3}\left(\fr{1}{\psi^{3}}+\t{x}_{2}^{3}+\t{x}_{3}^{3}-3\t{x}_{2}\t{x}_{3}\right)\ri) \nonum\\
  &=&-\int\fr{1}{W^{2}}x_{3}^{3}d\log x_{2}\wedge d\log x_{3}
   +\int\fr{\p}{\p x_{2}}\fr{x_{3}^{3}}{W}dx_{2}\wedge d\log x_{3} \nonum\\
  &&+\int\fr{\p}{\p x_{3}}\fr{x_{3}^{3}}{W}d\log x_{2}\wedge dx_{3}.
 \end{eqnarray}
A similar procedure using the transformation $\tilde{x}_{2}=x_{2}/\psi$ reproduces $\p/\p a_{2}$ and yields $x_{2}^{3}$,
 \begin{eqnarray}
  -\theta_{\psi}\int\fr{x_{3}^{3}}{W^{2}}d\log x_{2}\wedge d\log x_{3}
  &=&-2\int\fr{x_{2}^{3}x_{3}^{3}}{W^{3}}d\log x_{2}\wedge d\log x_{3}
   -\int\fr{\p}{\p x_{2}}\fr{x_{3}^{3}}{W^{2}}dx_{2}\wedge d\log x_{3}. \nonum\\
 \end{eqnarray}
Therefore we obtain
 \begin{eqnarray}
  \theta_{\psi}^2\int\Omega
  &=&-2\int\fr{\omega_{0}(x_{2}x_{3})^{2}}{W^{3}}
   +\int d\le(\fr{x_{3}^{3}}{W^{2}}\ri)\wedge(-d\log x_{3})\nonum\\
  &&+\theta_\psi\int d\le(\fr{x_{3}^{3}}{W}\ri)\wedge(d\log x_{3}-d\log x_{2})
   +\theta_{\psi}^{2}\int d\log W\wedge(-d\log x_{2}). 
 \end{eqnarray}
Note that from the usual derivative without coordinate transformation, we get
 \begin{eqnarray}
  \fr{1}{\psi}\theta_{\psi}\fr{1}{\psi}\theta_{\psi}\fr{1}{\psi}\int\Omega=2\int\fr{\omega_{0}}{W^{3}}(x_{2}x_{3})^{2}.
 \end{eqnarray}
From this we obtain the following differential equation 
 \begin{eqnarray}
  \fr{1}{\psi}\theta_{\psi}\fr{1}{\psi}\theta_{\psi}\fr{1}{\psi}\int\Omega+\theta_{\psi}^{2}\int\Omega
  &=&\int d\le(\fr{x_{3}^{3}}{W^{2}}\ri)\wedge(-d\log x_{3})
  +\theta_{\psi}\int d\le(\fr{x_{3}^{3}}{W}\ri)\wedge(d\log x_{3}-d\log x_{2})\nonum\\
  &&+\theta^{2}_{\psi}\int d\log W\wedge(-d\log x_{2}).
 \end{eqnarray}

\subsection{Double cubic}
Let us recall some fundamental formulas of $X_{3,3}[1^6]$ as follows: 
\begin{eqnarray}
W_{1}=\fr{1}{3}(x_{1}^{3}+x_{2}^{3}+x_{3}^{3})-\psi x_{4}x_{5}x_{6}, \ \ W_{2}=\fr{1}{3}(x_{4}^{3}+x_{5}^{3}+x_{6}^{3})-\psi x_{1}x_{2}x_{3}, \nonum\\
\Omega=\psi^{2}\int\fr{\omega_{0}}{W_{1}W_{2}}, \ \ \ 
\omega_{0}=\sum_{i=1}^{6}(-1)^{i}x_{i}dx_{1}\wedge\cdots\wedge\wh{dx_{i}}\wedge\cdots\wedge dx_{6}. 
\end{eqnarray}
In the following we mainly use the following two procedures: 
\begin{itemize}
\item I. 
Extract $x_i^3$ in the numerator by transforming as follows: 
transform $x_{i}=\t{x}_{i}/\psi$ for the term that contains $dx_{i}$ in $\omega_0$, 
and transform  $x_{j}=\psi \t{x}_{j}$ ($j\neq i$) for other terms. 
\item II. 
Extract $x_{1}x_{2}x_{3}$ in the numerator by transforming as follows: 
transform $x_{i}=\psi^{1/3}\t{x}_{i}$ ($i=1$, $2$, $3$) for the term that contains all of $dx_{1}$, $dx_{2}$, $dx_{3}$ in the $\omega_0$, 
and transform $x_{i}=\psi^{-1/3}\t{x}_{i}$ ($i=4$, $5$, $6$) for the term that contains all of $dx_{4}$, $dx_{5}$, $dx_{6}$ in the $\omega_0$. 

To extract $x_{4}x_{5}x_{6}$, we use the same procedure by exchanging ($i=1$, $2$, $3$) $\leftrightarrow$ ($i=4$, $5$, $6$). 
\end{itemize}
We use $\theta=\psi\p_{\psi}$ and the terms $B_{1.1}, \ \cdots$, are contributions from boundaries and 
the explicit formulas are listed in the later page. 

Firstly, by use of the procedure I, 
we consider the fourth order  derivative of $\Omega$ with respect to $\psi$. 
In the following we apply the procedure I to $x_{1}$, $x_{4}$, $x_{2}$, $x_{5}$ in turn. 
\begin{equation*}
\theta\fr{1}{\psi}\Omega
=\theta\psi\int\fr{\omega_{0}}{W_{1}W_{2}}
=\psi\int\fr{x_{1}^{3}}{W_{1}^{2}W_{2}}\omega_{0}
+\psi^{2}\int\fr{x_{4}x_{5}x_{6}}{W_{1}^{2}W_{2}}\omega_{0}
+B_{1.1}\label{1.5}
\end{equation*}
\begin{eqnarray}
  \theta\theta\fr{1}{\psi}\Omega
  &=&\theta\psi\int\fr{x_{1}^{3}}{W_{1}^{2}W_{2}}\omega_{0}
   +\theta\psi^{2}\int\fr{x_{4}x_{5}x_{6}}{W_{1}^{2}W_{2}}\omega_{0}
   +\theta B_{1.1}\nonum\\
  &=&\psi\int\fr{x_{1}^{3}x_{4}^{3}}{W_{1}^{2}W_{2}^{2}}\omega_{0}
   +\psi^{2}\int\fr{x_{1}^{4}x_{2}x_{3}}{W_{1}^{2}W_{2}^{2}}\omega_{0}
   +\theta\psi^{2}\int\fr{x_{4}x_{5}x_{6}}{W_{1}^{2}W_{2}}\omega_{0}
   +\theta B_{1.1}+B_{2.1}\label{1.6}\nonum
\end{eqnarray}
\begin{eqnarray}
  \theta\theta\theta\fr{1}{\psi}\Omega
  &=&\theta\psi\int\fr{x_{1}^{3}x_{4}^{3}}{W_{1}^{2}W_{2}^{2}}\omega_{0}
   +\theta\psi^{2}\int\fr{x_{1}^{4}x_{2}x_{3}}{W_{1}^{2}W_{2}^{2}}\omega_{0}
   +\theta\theta\psi^{2}\int\fr{x_{4}x_{5}x_{6}}{W_{1}^{2}W_{2}}\omega_{0}
   +\theta\theta B_{1.1}+\theta B_{2.1}\nonum\\
  &=&2\psi\int\fr{x_{1}^{3}x_{2}^{3}x_{4}^{3}}{W_{1}^{3}W_{2}^{2}}\omega_{0}
   +2\psi^{2}\int\fr{x_{1}^{3}x_{4}^{4}x_{5}x_{6}}{W_{1}^{3}W_{2}^{2}}\omega_{0}
   +\theta\psi^{2}\int\fr{x_{1}^{4}x_{2}x_{3}}{W_{1}^{2}W_{2}^{2}}\omega_{0}
   +\theta\theta\psi^{2}\int\fr{x_{4}x_{5}x_{6}}{W_{1}^{2}W_{2}}\omega_{0}\nonum\\
  &&+\theta\theta B_{1.1}+\theta B_{2.1}+B_{3.1}\nonum \label{1.7}
\end{eqnarray}
\begin{eqnarray}
  \theta\theta\theta\theta\fr{1}{\psi}\Omega
  &=&2\theta\psi\int\fr{x_{1}^{3}x_{2}^{3}x_{4}^{3}}{W_{1}^{3}W_{2}^{2}}\omega_{0}
   +2\theta\psi^{2}\int\fr{x_{1}^{3}x_{4}^{4}x_{5}x_{6}}{W_{1}^{3}W_{2}^{2}}\omega_{0}
   +\theta\theta\psi^{2}\int\fr{x_{1}^{4}x_{2}x_{3}}{W_{1}^{2}W_{2}^{2}}\omega_{0}
   +\theta\theta\theta\psi^{2}\int\fr{x_{4}x_{5}x_{6}}{W_{1}^{2}W_{2}}\omega_{0}\nonum\\
  &&+\theta\theta\theta B_{1.1}
   +\theta\theta B_{2.1}+\theta B_{3.1}\nonum\\
  &=&4\psi\int\fr{x_{1}^{3}x_{2}^{3}x_{4}^{3}x_{5}^{3}}{W_{1}^{3}W_{2}^{3}}\omega_{0}
   +4\psi^{2}\int\fr{x_{1}^{4}x_{2}^{4}x_{3}x_{4}^{3}}{W_{1}^{3}W_{2}^{3}}\omega_{0}
   +2\theta\psi^{2}\int\fr{x_{1}^{3}x_{4}^{4}x_{5}x_{6}}{W_{1}^{3}W_{2}^{2}}\omega_{0}
   +\theta\theta\psi^{2}\int\fr{x_{1}^{4}x_{2}x_{3}}{W_{1}^{2}W_{2}^{2}}\omega_{0}\nonum\\
  &&+\theta\theta\theta\psi^{2}\int\fr{x_{4}x_{5}x_{6}}{W_{1}^{2}W_{2}}\omega_{0}
   +\theta\theta\theta B_{1.1}
   +\theta\theta B_{2.1}+\theta B_{3.1}+B_{4.1}\label{1.8}
\end{eqnarray}

Secondly, we try to express each term of the r.h.s. of \eqref{1.8} as the $\psi$-derivative of $\Omega$. 
For the first term of the r.h.s. of \eqref{1.8}, 
by use of the procedure II, the following $\psi$-derivative formula can be obtained: 
\begin{eqnarray}
  4\psi\int\fr{(x_{1}x_{2}x_{4}x_{5})^{3}}{W_{1}^{3}W_{2}^{3}}\omega_{0}
  &=&\fr{1}{\psi}\theta\psi\int\fr{x_{1}^{3}x_{2}^{3}x_{4}^{2}x_{5}^{2}}{x_{6}W_{1}^{2}W_{2}^{3}}\omega_{0}
   -\fr{1}{3}\sum_{i=1}^{3}\int\fr{\p}{\p x_{i}}\fr{x_{i}x_{1}^{3}x_{2}^{3}x_{4}^{2}x_{5}^{2}}{x_{6}W_{1}^{2}W_{2}^{3}}\omega_{(123)}\nonum\\
  &+&\fr{1}{3}\sum_{i=4}^{6}\int\fr{\p}{\p x_{i}}\fr{x_{i}x_{1}^{3}x_{2}^{3}x_{4}^{2}x_{5}^{2}}{x_{6}W_{1}^{2}W_{2}^{3}}\omega_{(456)}, 
\end{eqnarray}
where $\omega_{(ijk)}$ expresses the part of $\omega_{0}$ which contains only $dx_{i},dx_{j},dx_{k}$. 

Next in order to extract $x_{3}^{3}$ in the numerator, 
we do the $\psi$-derivation after the next transformation: 
\begin{eqnarray}
&&\text{for \ } \omega_{(123)}: \ x_{1}=\psi^{1/3}\t{x}_{1},x_{2}=\psi^{1/3}\t{x}_{2},x_{3}=\psi^{-5/3}\t{x}_{3}, \nonum\\
&&\text{for \ } \omega_{(3456)}: \ x_{3}=\psi^{-2}\t{x}_{3},x_{4}=\psi^{-1/3}\t{x}_{4},x_{5}=\psi^{-1/3}\t{x}_{5},x_{6}=\psi^{-1/3}\t{x}_{6}, \nonum\\
&&\text{for \ } \omega_{(12456)}: \ x_{1}=\psi^{2}\t{x}_{1},x_{2}=\psi^{2}\t{x}_{2},x_{4}=\psi^{5/3}\t{x}_{4},x_{5}=\psi^{5/3}\t{x}_{5},x_{6}=\psi^{5/3}\t{x}_{6}, \nonum
\end{eqnarray}
\begin{equation}
=4\int\fr{x_{1}^{3}x_{2}^{3}x_{3}^{3}x_{4}^{2}x_{5}^{2}}{x_{6}W_{1}^{3}W_{2}^{3}}\omega_{0}+B_{0.1}\nonum
\end{equation}
Then we rewrite it as the $\psi$-derivative formula by using $\mathrm{II}$ again. 
\begin{eqnarray}
  &=&\fr{1}{\psi}\theta\int\fr{(x_{1}x_{2}x_{3}x_{4}x_{5})^{2}}{x_{6}W_{1}^{3}W_{2}^{2}}\omega_{0}
   +\fr{1}{3\psi}\sum_{i=1}^{3}\int\fr{\p}{\p x_{i}}\fr{x_{i}(x_{1}x_{2}x_{3}x_{4}x_{5})^{2}}{x_{6}W_{1}^{3}W_{2}^{2}}\omega_{(123)}\nonum\\
  &&-\fr{1}{3\psi}\sum_{i=4}^{6}\int\fr{\p}{\p x_{i}}\fr{x_{i}(x_{1}x_{2}x_{3}x_{4}x_{5})^{2}}{x_{6}W_{1}^{3}W_{2}^{2}}\omega_{(456)}
   +B_{0.1}\nonum
\end{eqnarray}
By use of the transformation as follows, we do the differentiation and extract $x_{6}^{3}$. 
\begin{eqnarray}
  &&\text{for \ }\omega_{(456)}: \ x_{4}=\psi^{1/3}\t{x}_{4},x_{5}=\psi^{1/3}\t{x}_{5},x_{6}=\psi^{-5/3}\t{x}_{6}, \nonum\\
  &&\text{for \ }\omega_{(1236)}: \ x_{1}=\psi^{-1/3}\t{x}_{1},x_{2}=\psi^{-1/3}\t{x}_{2},x_{3}=\psi^{-1/3}\t{x}_{3},x_{6}=\psi^{-2}\t{x}_{6}, \nonum\\
  &&\text{for \ }\omega_{(12345)}: \ x_{1}=\psi^{5/3}\t{x}_{1},x_{2}=\psi^{5/3}\t{x}_{2},x_{3}=\psi^{5/3}\t{x}_{3},x_{4}=\psi^{2}\t{x}_{4},x_{5}=\psi^{2}\t{x}_{5}, \nonum
\end{eqnarray}
\begin{equation}
=\fr{4}{\psi}\int\fr{(x_{1}x_{2}x_{3}x_{4}x_{5}x_{6})^{2}}{W_{1}^{3}W_{2}^{3}}\omega_{0}+B_{0.1}+B_{0.2}\nonum
\end{equation}
We can rewrite this as the forth derivative of $\Omega$ with respect to $\psi$ by using transformations II. 
\begin{eqnarray}
  &=&\fr{1}{16\psi^{3}}\theta\fr{1}{\psi^{2}}\theta\psi^{2}\theta\fr{1}{\psi^{2}}\theta\fr{1}{\psi^{2}}\Omega
   -\fr{1}{16\psi^{3}}\biggl(\theta\fr{1}{\psi^{2}}\theta\psi^{2}\theta\fr{1}{\psi^{2}}B_{1}
   +\theta\fr{1}{\psi^{2}}\theta\psi^{2}B_{2}
   +\theta\fr{1}{\psi^{2}}B_{3}+B_{4}\biggr)\nonum\\
  &&+B_{0.1}+B_{0.2}. 
\end{eqnarray}
The fifth term can be written as the first derivative of $\Omega$ by using transformation II: 
\begin{equation}
  \psi^{2}\int\fr{x_{4}x_{5}x_{6}}{W_{1}^{2}W_{2}}\omega_{0}
  =\frac{\psi}{2}\theta\frac{1}{\psi^{2}}\Omega-B_{1.2}. 
\end{equation}
The fourth term can be expressed as follows 
by use of the formula which is obtained by differentiating of \eqref{1.5} with respect to $\fr{\psi}{2}\theta\fr{1}{\psi}$ and transforming as II: 
\begin{eqnarray}
  \psi^{2}\int\fr{x_{1}^{4}x_{2}x_{3}}{W_{1}^{2}W_{2}^{2}}\omega_{0}
  &=&\fr{\psi}{2}\theta\fr{1}{\psi}\theta\fr{1}{\psi}\Omega-\fr{\psi}{4}\theta\theta\fr{1}{\psi^{2}}\Omega
   -\fr{\psi}{2}\theta\fr{1}{\psi}(B_{1.1}-B_{1.2})-\fr{\psi}{2}B_{1.3}. \nonum
\end{eqnarray}
The third term can be expressed as follows 
by use of the formula which is obtained by differentiating of \eqref{1.6} with respect to $\fr{\psi}{4}\theta\fr{1}{\psi}$ and transforming as II: 
\begin{eqnarray}
  \psi^{2}\int\fr{x_{1}^{3}x_{4}^{4}x_{5}x_{6}}{W_{1}^{3}W_{2}^{2}}\omega_{0}
  &=&\fr{\psi}{4}\theta\fr{1}{\psi}\theta\theta\fr{1}{\psi}\Omega
   -\fr{\psi}{8}\theta\theta\fr{1}{\psi}\theta\fr{1}{\psi}\Omega
   +\fr{\psi}{16}\theta\theta\theta\fr{1}{\psi^{2}}\Omega
   +\fr{\psi}{8}\theta\theta\fr{1}{\psi}(B_{1.1}-B_{1.2})
   +\fr{\psi}{4}\theta\fr{1}{2}B_{1.3}\nonum\\
  &&-\fr{\psi}{4}\theta\fr{1}{\psi}\theta\fr{\psi}{2}\theta\fr{1}{\psi^{2}}\Omega
   -\fr{\psi}{4}\theta\fr{1}{\psi}\theta (B_{1.1}-B_{1.2})
   -\fr{\psi}{4}\theta\fr{1}{\psi} B_{2.1}-\fr{\psi}{4}B_{2.2}. \nonum
\end{eqnarray}
The second term can be expressed as follows 
by use of the formula which is obtained by differentiating of \eqref{1.7} with respect to $\psi\theta\fr{1}{2\psi}$ 
and transforming as II: 
\begin{eqnarray}
  4\psi^{2}\int\fr{x_{1}^{4}x_{2}^{4}x_{3}x_{4}^{3}}{W_{1}^{3}W_{2}^{3}}\omega_{0}
  &=&\psi\theta\fr{1}{2\psi}\theta\theta\theta\fr{1}{\psi}\Omega
   -2\psi\theta\fr{1}{2\psi}\biggl[\fr{\psi}{4}\theta\fr{1}{\psi}\theta\theta\fr{1}{\psi}\Omega
   -\fr{\psi}{8}\theta\theta\fr{1}{\psi}\theta\fr{1}{\psi}\Omega
   +\fr{\psi}{16}\theta\theta\theta\fr{1}{\psi^{2}}\Omega \nonum\\
  &&+\fr{\psi}{8}\theta\theta\fr{1}{\psi}(B_{1.1}-B_{1.2})
   +\fr{\psi}{4}\theta\fr{1}{2}B_{1.3}
   -\fr{\psi}{4}\theta\fr{1}{\psi}\theta\fr{\psi}{2}\theta\fr{1}{\psi^{2}}\Omega\nonum\\
  &&-\fr{\psi}{4}\theta\fr{1}{\psi}\theta (B_{1.1}-B_{1.2})
   -\fr{\psi}{4}\theta\fr{1}{\psi}B_{2.1}-\fr{\psi}{4}B_{2.2}\biggr]\nonum\\
  &&-\psi\theta\fr{1}{2\psi}\theta\le[\fr{\psi}{2}\theta\fr{1}{\psi}\theta\fr{1}{\psi}\Omega-\fr{\psi}{4}\theta\theta\fr{1}{\psi^{2}}\Omega
   -\fr{\psi}{2}\theta\fr{1}{\psi}(B_{1.1}-B_{1.2})-\fr{\psi}{2}B_{1.3}\ri]\nonum\\
  &&-\psi\theta\fr{1}{2\psi}\theta\theta\fr{\psi}{2}\theta\fr{1}{\psi^{2}}\Omega
   -\psi\theta\fr{1}{2\psi}\theta\theta (B_{1.1}-B_{1.2})
   -\psi\theta\fr{1}{2\psi}\theta B_{2.1}
   -\psi\theta\fr{1}{2\psi}B_{3.1}-\psi B_{3.2}. \nonum
\end{eqnarray}
Therefore we lead to the following deferential formula: 

\begin{eqnarray}
  \theta\theta\theta\theta\fr{1}{\psi}\Omega
  &=&\fr{1}{16\psi^{3}}\theta\fr{1}{\psi^{2}}\theta\psi^{2}\theta\fr{1}{\psi^{2}}\theta\fr{1}{\psi^{2}}\Omega
   +B_{0.1}+B_{0.2}
   +\psi\theta\fr{1}{2\psi}\theta\theta\theta\fr{1}{\psi}\Omega\nonum\\
  &&-2\psi\theta\fr{1}{2\psi}\biggl[\fr{\psi}{4}\theta\fr{1}{\psi}\theta\theta\fr{1}{\psi}\Omega
   -\fr{\psi}{8}\theta\theta\fr{1}{\psi}\theta\fr{1}{\psi}\Omega
   +\fr{\psi}{16}\theta\theta\theta\fr{1}{\psi^{2}}\Omega\nonum\\
  &&+\fr{\psi}{8}\theta\theta\fr{1}{\psi}B_{1.1}
   -\fr{\psi}{4}\theta\fr{1}{\psi}\theta\fr{\psi}{2}\theta\fr{1}{\psi^{2}}\Omega
   -\fr{\psi}{4}\theta\fr{1}{\psi}\theta B_{1.1}
   -\fr{\psi}{4}\theta\fr{1}{\psi}B_{2.1}\biggr]\nonum\\
  &&-\psi\theta\fr{1}{2\psi}\theta\le[\fr{\psi}{2}\theta\fr{1}{\psi}\theta\fr{1}{\psi}\Omega-\fr{\psi}{4}\theta\theta\fr{1}{\psi^{2}}\Omega
   -\fr{\psi}{2}\theta\fr{1}{\psi}B_{1.1}-\fr{\psi}{2}B_{1.3}\ri]\nonum\\
  &&-\psi\theta\fr{1}{2\psi}\theta\theta\fr{\psi}{2}\theta\fr{1}{\psi^{2}}\Omega
   -\psi\theta\fr{1}{2\psi}\theta\theta B_{1.1}
   -\psi\theta\fr{1}{2\psi}\theta B_{2.1}
   +\psi\theta\fr{1}{2\psi}B_{3.1}\nonum\\
  &&+2\theta\biggl(\fr{\psi}{4}\theta\fr{1}{\psi}\theta\theta\fr{1}{\psi}\Omega
   -\fr{\psi}{8}\theta\theta\fr{1}{\psi}\theta\fr{1}{\psi}\Omega
   +\fr{\psi}{16}\theta\theta\theta\fr{1}{\psi^{2}}\Omega\nonum\\
  &&+\fr{\psi}{8}\theta\theta\fr{1}{\psi}B_{1.1}
   -\fr{\psi}{4}\theta\fr{1}{\psi}\theta\fr{\psi}{2}\theta\fr{1}{\psi^{2}}\Omega
   -\fr{\psi}{4}\theta\fr{1}{\psi}\theta B_{1.1}
   -\fr{\psi}{4}\theta\fr{1}{\psi}B_{2.1}\biggr)\nonum\\
  &&+\theta\theta\le(\fr{\psi}{2}\theta\fr{1}{\psi}\theta\fr{1}{\psi}\Omega-\fr{\psi}{4}\theta\theta\fr{1}{\psi^{2}}\Omega
   -\fr{\psi}{2}\theta\fr{1}{\psi}B_{1.1}-\fr{\psi}{2}B_{1.3}\ri)\nonum\\
  &&+\theta\theta\theta\fr{\psi}{2}\theta\fr{1}{\psi^{2}}\Omega+\theta\theta\theta B_{1.1}
   +\theta\theta B_{2.1}+\theta B_{3.1}+B_{4.1}. 
\end{eqnarray}
\begin{eqnarray*}
B_{1.1}&=&\psi\int d\le(\fr{x_{1}}{W_{1}W_{2}}\omega_{1}\ri), \ \ 
B_{1.2}=\fr{\psi}{6}\sum_{i=1}^{3}\int\fr{\p}{\p x_{i}}\fr{x_{i}}{W_{1}W_{2}}\omega_{(123)}
       -\fr{\psi}{6}\sum_{i=4}^{6}\int\fr{\p}{\p x_{i}}\fr{x_{i}}{W_{1}W_{2}}\omega_{(456)}\\
B_{1.3}&=&-\fr{1}{3}\sum_{i=1}^{3}\int\fr{\p}{\p x_{i}}\fr{x_{i}x_{1}^{3}}{W_{1}^{2}W_{2}}\omega_{(123)}
        +\fr{1}{3}\sum_{i=4}^{6}\int\fr{\p}{\p x_{i}}\fr{x_{i}x_{1}^{3}}{W_{1}^{2}W_{2}}\omega_{(456)}\\
B_{2.1}&=&\psi\int d\le(\fr{x_{1}^{3}x_{4}}{W_{1}^{2}W_{2}}\omega_{4}\ri), \ \ 
B_{2.2}=\fr{1}{3}\sum_{i=1}^{3}\int\fr{\p}{\p x_{i}}\fr{x_{i}x_{1}^{3}x_{4}^{3}}{W_{1}^{2}W_{2}^{2}}\omega_{(123)}
        -\fr{1}{3}\sum_{i=4}^{6}\int\fr{\p}{\p x_{i}}\fr{x_{i}x_{1}^{3}x_{4}^{3}}{W_{1}^{2}W_{2}^{2}}\omega_{(456)}\\
B_{3.1}&=&\psi\int d\le(\fr{x_{1}^{3}x_{4}^{3}x_{2}}{W_{1}^{2}W_{2}^{2}}\omega_{2}\ri), \ \ 
B_{3.2}=-\fr{1}{3}\sum_{i=1}^{3}\int\fr{\p}{\p x_{i}}\fr{x_{i}x_{1}^{3}x_{2}^{3}x_{4}^{3}}{W_{1}^{3}W_{2}^{2}}\omega_{(123)}
        +\fr{1}{3}\sum_{i=4}^{6}\int\fr{\p}{\p x_{i}}\fr{x_{i}x_{1}^{3}x_{2}^{3}x_{4}^{3}}{W_{1}^{3}W_{2}^{2}}\omega_{(456)}\\
B_{4.1}&=&2\psi\int d\le(\fr{x_{1}^{3}x_{2}^{3}x_{4}^{3}x_{5}}{W_{1}^{3}W_{2}^{2}}\omega_{5}\ri)
\end{eqnarray*}
\begin{eqnarray*}
  B_{0.1}&=&-2\sum_{i=1}^{2}\int\fr{\p}{\p x_{i}}\fr{x_{i}x_{1}^{3}x_{2}^{3}x_{4}^{2}x_{5}^{2}}{x_{6}W_{1}^{2}W_{2}^{3}}\omega_{(12456)}
   -\fr{5}{3}\sum_{i=4}^{6}\int\fr{\p}{\p x_{i}}\fr{x_{i}x_{1}^{3}x_{2}^{3}x_{4}^{2}x_{5}^{2}}{x_{6}W_{1}^{2}W_{2}^{3}}\omega_{(12456)}\\
  &&-\fr{1}{3}\sum_{i=1}^{2}\int\fr{\p}{\p x_{i}}\fr{x_{i}x_{1}^{3}x_{2}^{3}x_{4}^{2}x_{5}^{2}}{x_{6}W_{1}^{2}W_{2}^{3}}\omega_{(123)}
   +\fr{5}{3}\int\fr{\p}{\p x_{3}}\fr{x_{3}x_{1}^{3}x_{2}^{3}x_{4}^{2}x_{5}^{2}}{x_{6}W_{1}^{2}W_{2}^{3}}\omega_{(123)}\\
  &&+\fr{1}{3}\sum_{i=4}^{6}\int\fr{\p}{\p x_{i}}\fr{x_{i}x_{1}^{3}x_{2}^{3}x_{4}^{2}x_{5}^{2}}{x_{6}W_{1}^{2}W_{2}^{3}}\omega_{(3456)}
   +2\int\fr{\p}{\p x_{3}}\fr{x_{3}x_{1}^{3}x_{2}^{3}x_{4}^{2}x_{5}^{2}}{x_{6}W_{1}^{2}W_{2}^{3}}\omega_{(3456)}\\
  &&-\fr{1}{3}\sum_{i=1}^{3}\int\fr{\p}{\p x_{i}}\fr{x_{i}x_{1}^{3}x_{2}^{3}x_{4}^{2}x_{5}^{2}}{x_{6}W_{1}^{2}W_{2}^{3}}\omega_{(123)}
   +\fr{1}{3}\sum_{i=4}^{6}\int\fr{\p}{\p x_{i}}\fr{x_{i}x_{1}^{3}x_{2}^{3}x_{4}^{2}x_{5}^{2}}{x_{6}W_{1}^{2}W_{2}^{3}}\omega_{(456)}\\
  B_{0.2}&=&\fr{1}{3\psi}\sum_{i=1}^{3}\int\fr{\p}{\p x_{i}}\fr{x_{i}(x_{1}x_{2}x_{3}x_{4}x_{5})^{2}}{x_{6}W_{1}^{3}W_{2}^{2}}\omega_{(123)}
   -\fr{1}{3\psi}\sum_{i=4}^{6}\int\fr{\p}{\p x_{i}}\fr{x_{i}(x_{1}x_{2}x_{3}x_{4}x_{5})^{2}}{x_{6}W_{1}^{3}W_{2}^{2}}\omega_{(456)}\\
  &&-\fr{2}{\psi}\sum_{i=4}^{5}\int\fr{\p}{\p x_{i}}\fr{x_{i}(x_{1}x_{2}x_{3}x_{4}x_{5})^{2}}{x_{6}W_{1}^{3}W_{2}^{2}}\omega_{12345}
   -\fr{5}{3\psi}\sum_{i=1}^{3}\int\fr{\p}{\p x_{i}}\fr{x_{i}(x_{1}x_{2}x_{3}x_{4}x_{5})^{2}}{x_{6}W_{1}^{3}W_{2}^{2}}\omega_{12345}\\
  &&+\fr{1}{3\psi}\sum_{i=1}^{3}\int\fr{\p}{\p x_{i}}\fr{x_{i}(x_{1}x_{2}x_{3}x_{4}x_{5})^{2}}{x_{6}W_{1}^{3}W_{2}^{2}}\omega_{(1236)}
   +\fr{2}{\psi}\int\fr{\p}{\p x_{6}}\fr{x_{6}(x_{1}x_{2}x_{3}x_{4}x_{5})^{2}}{x_{6}W_{1}^{3}W_{2}^{2}}\omega_{(1236)}\\
  &&-\fr{1}{3\psi}\sum_{i=4}^{5}\int\fr{\p}{\p x_{i}}\fr{x_{i}(x_{1}x_{2}x_{3}x_{4}x_{5})^{2}}{x_{6}W_{1}^{3}W_{2}^{2}}\omega_{(456)}
   +\fr{5}{3\psi}\int\fr{\p}{\p x_{6}}\fr{x_{6}(x_{1}x_{2}x_{3}x_{4}x_{5})^{2}}{x_{6}W_{1}^{3}W_{2}^{2}}\omega_{(456)}\\
B_{1}&=&-\fr{1}{3}\sum_{i=1}^{3}\int\fr{\p}{\p x_{i}}\fr{x_{i}}{W_{1}W_{2}}\omega_{(123)}
      +\fr{1}{3}\sum_{i=4}^{6}\int\fr{\p}{\p x_{i}}\fr{x_{i}}{W_{1}W_{2}}\omega_{(456)}\\
B_{2}&=&-\fr{2}{3\psi}\sum_{i=1}^{3}\int\fr{\p}{\p x_{i}}\fr{x_{i}x_{1}x_{2}x_{3}}{W_{1}W_{2}^{2}}\omega_{(123)}
      +\fr{3}{3\psi}\sum_{i=4}^{6}\int\fr{\p}{\p x_{i}}\fr{x_{i}x_{1}x_{2}x_{3}}{W_{1}W_{2}^{2}}\omega_{(456)}\\
B_{3}&=&\fr{8}{3}\psi^{2}\sum_{i=1}^{3}\int\fr{\p}{\p x_{i}}\fr{x_{i}(x_{1}x_{2}x_{3})^{2}}{W_{1}W_{2}^{3}}\omega_{(123)}
      -\fr{8}{3}\psi^{2}\sum_{i=4}^{6}\int\fr{\p}{\p x_{i}}\fr{x_{i}(x_{1}x_{2}x_{3})^{2}}{W_{1}W_{2}^{3}}\omega_{(456)}\\
B_{4}&=&\fr{16}{3}\psi^{2}\sum_{i=1}^{3}\int\fr{\p}{\p x_{i}}\fr{x_{i}(x_{1}x_{2}x_{3})^{2}x_{4}x_{5}x_{6}}{W_{1}W_{2}^{3}}\omega_{(123)}
      -\fr{16}{3}\psi^{2}\sum_{i=4}^{6}\int\fr{\p}{\p x_{i}}\fr{x_{i}(x_{1}x_{2}x_{3})^{2}x_{4}x_{5}x_{6}}{W_{1}W_{2}^{3}}\omega_{(456)}
\end{eqnarray*}
So we lead to the following differential operator, 
\begin{equation}
  \mathcal{L}
  =\fr{1}{16}\le(\psi^{3}-\fr{1}{\psi^{3}}\ri)\p_{\psi}^{4}
  +\fr{3}{8}\le(\psi^{2}+\fr{1}{\psi^{4}}\ri)\p_{\psi}^{3}
  +\fr{1}{16}\le(7\psi-\fr{23}{\psi^{5}}\ri)\p_{\psi}^{2}
  +\fr{1}{16}\le(1+\fr{55}{\psi^{6}}\ri)\p_{\psi}
  -\fr{4}{\psi^{7}},\\ \label{PF_111} 
\end{equation}
and the boundary contributions which give the inhomogeneous term of the differential equation. 

Now we turn to evaluate the inhomogeneous term.
\begin{equation}
  \mathcal{L}\int_{\Gamma}\Omega=
  \mathcal{L}\psi^{2}\int_{T_{\epsilon}(\Gamma)}\fr{\omega_{0}}{W_{1}W_{2}}=\int_{T_{\epsilon}(C_{+}-C_{-})}\beta. 
\end{equation}
We introduce the derivative $\theta=z\p_{z}$ $(z=(3\psi)^{-6})$, 
to simplify the equation \eqref{PF_111}.  The standard form of the Picard-Fuchs 
differential operator $\mathcal{L}_{PF}$ is given by 
\begin{equation}
  \mathcal{L}_{PF}=\theta^{4}-9z(3\theta+1)^{2}(3\theta+2)^{2}.
\end{equation} 
${\cal L}_{PF}$ is related to ${\cal L}$ by
\begin{equation}
  \mathcal{L}_{PF}=\fr{\psi}{81}\mathcal{L}. 
\end{equation}
We need to know explicit form of the defining equations of the boundary curves. 
As noted above the curves are defined as the intersection of hyperplanes 
$P=\{x_{1}+x_{2}=0, x_{4}+x_{5}=0\}$ \eqref{double cubic curve def} and
rewritten by 
\begin{equation}
  C_{\pm}=\le\{x_{1}+x_{2}=0,\quad x_{4}+x_{5}=0,\quad x_{3}^{3}+3\psi x_{4}^{2}x_{6}=0,\quad x_{1}=\pm\fr{x_{3}^{4}}{(3\psi)^{2}x_{4}^{3}}\ri\}. \label{111_111_C}
\end{equation}
In addition, there are two intersection points $C_\pm$ such as
$p_{1}=\{x_{1}=x_{2}=x_{3}=x_{6}=x_{4}+x_{5}=0\}$ and 
$p_{2}=\{x_{3}=x_{4}=x_{5}=x_{6}=x_{1}+x_{2}=0\}$. 
Since we can apply the tubes $T_{\epsilon}(C_{\pm})$ into $P$ except for the neighborhoods of $p_{1}$ and $p_{2}$, 
our evaluation of the inhomogeneous term of the Picard-Fuchs equation as integration of the exact term $d\beta$
on the tubes $T_{\epsilon}(C_{\pm})$ is localized around these points \cite{MW}. 

We choose locally resolved coordinates around the point $p_{1}$ in the $x_5=1$ patch. 
\begin{equation}
  X=\fr{x_{2}^{2}}{x_{1}x_{3}^{4}},\quad Z=\fr{x_{1}^{2}}{x_{2}x_{3}^{4}},\quad Y=x_{3}^{3},\quad T=x_{4}, 
  \quad 1=x_{5},\quad U=\fr{x_{6}}{x_{3}^{3}}. 
\end{equation}
In terms of these coordinates, the boundary $C_{\pm}$ is parametrized as
\begin{equation}
T=-1,\quad U=-\fr{1}{3\psi}, \ X(=-Z)=\mp\fr{1}{(3\psi)^{2}}, \ Y=re^{i\varphi} \ (r>0, \ 0\leq\varphi<2\pi). 
\end{equation}
The singular point $p_1$ is resolved and splits into two points  $p_{1,\pm}$. 
The tube around $p_{1,+}$, $T_{\epsilon}(C_{+};p_{1,+})$ is 
generated by a vector $v$: 
\begin{equation}
  v=\fr{f(r)}{r}e^{i\xi}e^{i\chi}\p_{T}+\fr{\alpha}{\psi}e^{-3i\varphi}e^{i\chi}\p_{X}
   -\fr{\alpha}{\psi}e^{-3i\varphi}e^{i\chi}\p_{Z}+e^{i\xi}e^{i\chi}\p_{U}, 
\end{equation}
\begin{equation}
T=-1+\epsilon \fr{f(r)}{r}e^{i\chi}e^{i\xi}, \ 
X=-Z=-\fr{1}{(3\psi)^{2}}+\epsilon \fr{\alpha}{\psi}e^{-3i\varphi}e^{i\chi}, \ 
U=-\fr{1}{3\psi}+\epsilon e^{i\chi}e^{i\xi}, \ 
Y=re^{i\varphi}, \nonumber
\end{equation}
\begin{equation}
(0\leq\varphi\leq 2\pi, \ 0\leq\chi\leq 2\pi, \ 0\leq\xi\leq 2\pi). \label{111_cordinates}
\end{equation}
\begin{eqnarray}
  dx_{1}dx_{3}dx_{4}dx_{6}&=&\fr{1}{3}Y^{5/3}dXdYdTdU\nonumber\\
  &=&-\fr{i\alpha\epsilon^{3}}{3\psi}r^{5/3}e^{-i\varphi/3}e^{3i\chi}e^{2i\xi}\le(f^{\pr}(r)-\fr{f(r)}{r}\ri)drd\varphi d\chi d\xi. 
\end{eqnarray}
This tube satisfies the conditions that ensure not to intersect the other boundary: 
\begin{eqnarray}
  &&d_{v}W_{1}|_{C_{+}}=e^{\chi}e^{i\varphi}e^{i\xi}\le(\fr{1}{3}f(r)+\psi r\ri)\ne 0, \quad (\psi>0)\\
  &&d_{v}W_{2}|_{C_{+}}=e^{\chi}\le[\fr{f(r)}{r}e^{i\xi}+\fr{r^{3}}{(3\psi)^{2}}(2\alpha-e^{i\xi}e^{3i\varphi})\ri]\ne 0. \quad (\alpha>1) 
\end{eqnarray}
The inhomogeneous term from $T_{\epsilon}(C_{+};p_{1})$ is 
\begin{eqnarray}
  \int_{T_{\epsilon}(C_{+};p_{1,+})}\beta
  &=&\biggl(-2\psi\theta\fr{1}{2\psi}\fr{\psi}{8}\theta\theta\fr{1}{\psi}
   +2\psi\theta\fr{1}{2\psi}\fr{\psi}{4}\theta\fr{1}{\psi}\theta 
   +\psi\theta\fr{1}{2\psi}\theta\fr{\psi}{2}\theta\fr{1}{\psi}
  -\psi\theta\fr{1}{2\psi}\theta\theta\nonumber\\
  &&+2\theta\fr{\psi}{8}\theta\theta\fr{1}{\psi}
   -2\theta\fr{\psi}{4}\theta\fr{1}{\psi}\theta-\theta\theta\fr{\psi}{2}\theta\fr{1}{\psi}+\theta\theta\theta\biggr)
   \psi\int\fr{x_{1}}{W_{1}W_{2}}\omega_{1}\nonumber\\
  &&+\le(\psi\theta\fr{1}{2\psi}+\theta\ri)\psi\int \fr{x_{1}^{3}x_{4}^{3}x_{2}}{W_{1}^{2}W_{2}^{2}}\omega_{2}.
\end{eqnarray}
 The non-zero contribution to the integral on $T_{\epsilon}(C_{+};p_{1,+})$ yields from
\begin{eqnarray}
  \le(\fr{1}{2}\psi\theta_{\psi}+\theta_{\psi}\psi\ri)\int_{T_{\epsilon}(C_{+};p_{1,+})}\fr{x_{1}^{3}x_{4}^{3}x_{2}}{W_{1}^{2}W_{2}^{2}}dx_{1}dx_{3}dx_{4}dx_{6}
  =\fr{2i\pi^{3}}{3\psi^{5}}.
\end{eqnarray}
In a similar way, we find the non-zero contribution from $C_{-}$ is 
\begin{eqnarray}
  \le(\fr{1}{2}\psi\theta_{\psi}+\theta_{\psi}\psi\ri)\int_{T_{\epsilon}(C_{-};p_{1,-})}\fr{x_{1}^{3}x_{4}^{3}x_{2}}{W_{1}^{2}W_{2}^{2}}dx_{1}dx_{3}dx_{4}dx_{6}
  =-\fr{2i\pi^{3}}{3\psi^{5}}.
\end{eqnarray}
The definition of local coordinates around $p_{2}$ is given by just exchanging $x_{1} \leftrightarrow x_{4}$, $x_{2} \leftrightarrow x_{5}$ and $x_{3} \leftrightarrow x_{6}$ 
in \eqref{111_cordinates}. 
However, we find there is no contribution from the integrations around $p_{2}$. 

Note that another possibility of contribution to the inhomogeneous term, 
which originates from the action of the differential operator on the three-chain, 
has no contribution like all other known models. 
We refer \cite{MW} for detail. 
As a result of computations, we finally find \eqref{111_result}. 

\newpage

\end{document}